\documentclass{aa}

\usepackage{graphicx}
\usepackage{xcolor}
\usepackage{colortbl}
\usepackage{txfonts}
\usepackage{pifont}
\usepackage[colorlinks=true,allcolors=blue]{hyperref}
\usepackage{parskip}
\usepackage{multirow}

\usepackage[font=small,labelfont=bf,tableposition=top]{caption}
\DeclareCaptionLabelFormat{andtable}{#1~#2  \&  \tablename~\thetable}

\begin{document}

   \title{The quiescent population at $0.5\le z \le 0.9$: Environmental impact on the mass-size relation}

   \author{M. Figueira\inst{\ref{inst:MPIfR}, \ref{inst:NCBJ}}
      \and M. Siudek\inst{\ref{inst:ISS}}
      \and A. Pollo\inst{\ref{inst:NCBJ}, \ref{inst:Jag}}
      \and J. Krywult \inst{\ref{inst:IPKielce}}
      \and D. Vergani \inst{\ref{inst:INAF_bologna}}
      \and M. Bolzonella\inst{\ref{inst:INAF_bologna}}
      \and O. Cucciati\inst{\ref{inst:INAF_bologna}}
      \and A.~Iovino\inst{\ref{inst:INAF_brera}}
          }

   \institute{Max-Planck-Institut für Radioastronomie, Auf dem Hügel 69, 53121, Bonn, Germany\label{inst:MPIfR}
         \and National Centre for Nuclear Research, Pasteura 7, 02-093 Warszawa, Poland\label{inst:NCBJ}
         \and Institute of Space Sciences (ICE, CSIC), Campus UAB, Carrer de Can Magrans, s/n, 08193 Barcelona, Spain\label{inst:ISS}
         \and Astronomical Observatory of the Jagiellonian University, Orla 171, 30-244 Kraków, Poland\label{inst:Jag}
         \and Institute of Physics, Jan Kochanowski University, ul. Swi\k{e}tokrzyska 15, 25-406 Kielce, Poland\label{inst:IPKielce}      
         \and INAF - Osservatorio di Astrofisica e Scienza dello Spazio, Via Piero Gobetti 93/3, I-40129 Bologna, Italy\label{inst:INAF_bologna}
         \and INAF - Osservatorio Astronomico di Brera, Via Brera 28, 20122 Milano, via E. Bianchi 46, 23807 Merate, Italy\label{inst:INAF_brera}
            }
   \date{Received 22/08/2023; accepted 02/04/2024}

 
  \abstract
   {How the quiescent galaxies evolve with redshift and the factors that impact their evolution are still debated. It is still unclear what the dominant mechanisms of passive galaxy growth are and what role is played by the environment in shaping their evolutionary paths over cosmic time.}
   {The population of quiescent galaxies is altered over time by several processes that can affect their mean properties. Our aim is to study the mass-size relation (MSR) of the quiescent population and to understand how the environment shapes the MSR at intermediate redshift. 
   }
   {We used the VIMOS Public Extragalactic Redshift Survey (VIPERS), a large spectroscopic survey of $\sim$90~000 galaxies in the redshift range $0.5\le z \le 1.2$. We selected a mass-complete sample of 4786 passive galaxies based on the NUVrK diagram and refined it using the $D_n4000$ spectral index to study the MSR of the passive population over $0.5\le z \le 0.9$. The impact of the environment on the MSR and on the growth of the quiescent population is studied through the density contrast.}
   {The slope and the intercept of the MSR, $\alpha=0.62\pm 0.04$ and $\textrm{log}(A)=0.52\pm 0.01$, agree well with values from the literature at the same redshift. The intercept decreases with redshift, $R_e(z)=8.20\times (1+z)^{-1.70}$, while the slope remains roughly constant, and the same trend is observed in the low-density (LD) and high-density (HD) environments. Thanks to the largest spectroscopic sample at $0.5\le z \le 0.9$, these results are not prone to redshift uncertainties from photometric measurements. We find that the average size of the quiescent population in the LD and HD environments are identical within $3\sigma$ and this result is robust against a change in the definition of the LD and HD environments or a change in the selection of quiescent galaxies. In the LD and HD environments, $\sim$30 and $\sim$40\% of the population have experienced a minor merger process between $0.5\le z \le 0.9$. However, minor mergers account only for 30 to 40\% of the size evolution in this redshift range, the remaining evolution likely being due to the progenitor bias.}
   {}


   \maketitle
%

\section{Introduction}

The formation and evolution of galaxies led to a wide variety of morphology that are observed from the high redshift down to the local Universe: elliptical, lenticular, spiral, barred spiral, and irregular. The first evolutionary model, where elliptical or early-type galaxies (ETGs) were thought to evolve into spiral or late-type galaxies (LTGs), was proposed by \citet{hub36}, and was known as the tuning fork diagram. Throughout the increased efforts that were made during the past decades in that field, it has nonetheless become clear that the evolutionary picture is far more complicated.\\ These advances were made possible thanks to the multiple statistical photometric and spectroscopic surveys from the local Universe and beyond that were performed in the last two decades. Among them are the 2dF Galaxy Redshift Survey (2dFGRS, $z\sim0.1$, \citealt{col01,col03}), the Sloan Digital Sky Survey (SDSS, $z\le 0.3$, \citealt{yor00, ahn14}), the VIMOS public extragalactic redshift survey (VIPERS, $0.4\le z\le 1.2$, \citealt{guz13,sco18}), the Galaxy And Mass Assembly survey (GAMA, $z<0.5$, \citealt{dri16}), or the Deep Extragalactic VIsible Legacy Survey (DEVILS, $0.3 < z < 1.0$, \citealt{dav18}). Thrilling new results and breakthroughs are also expected from the \textit{James Webb Space telescope} (\textit{JWST}) Advanced Deep Extragalactic Survey (JADES, $2\le z\le 11$, \citealt{eis23}) as well as from future observations with \textit{Euclid} \citep{EuclidcollabI} and the Rubin Observatory Legacy Survey of Space and Time (LSST).

One of the most interesting results unveiled in the field of galaxy evolution is the fact that massive metal-rich ETGs ($\ge10^{11}$~$M_{\odot}$) are more compact at high redshift ($z\ge 2$) compared to the local Universe, with a difference in stellar density reaching two orders of magnitude (e.g., \citealt{dad05,tru06a,tru06b,zir07,van08,cim08,dok08}). At first sight, it appears that ETGs have to undergo a substantial size evolution from $z\sim 2-3$ ($\sim10-11$~Gyr ago) through different mechanisms (e.g., \citealt{dok05, tru06a,tru06b, fra08, van08, dok08, dam11, cim12, ham22}) in order to match observations performed in the local Universe. However, the size evolution of ETGs is still debated and might not be required to explain the mean properties of this population at $z=0$. For instance, \citet{dam09} could not find a satisfactory size evolution mechanism at $z\sim1.5$, and \citet{pog13} find only a mild size increase from $z\sim 0.7$ to $z\sim 0.05$. The number density evolution also seems to be inconsistent with a strong size evolution \citep{pog10}, the number density of compact ETGs appears to be in agreement with local Universe observations in clusters \citep{sar10}, or to slightly decrease \citep{gar16,gar17}. In addition, the numbers of compact ETGs could be underestimated in SDSS \citep{dam09,dam11}, leading to an apparent decrease in the number density of those compact ETGs.\\

The change in size and mass that compact galaxies might experience over cosmic time is imprinted on a fundamental plane known as the mass-size relation (MSR) that is often used to quantify the evolution of galaxies, their assembly, or the properties of their DM haloes (e.g., \citealt{she03,cim12,ich12,hue13a,van14,fai17,fav18,jia18,mow19b,mow19a,dam19,bar22,afa23}). Several works have already shown that LTGs and ETGs are following their own MSR (e.g., \citealt{she03,ich12,van14,fai17,mow19b,mos20,ned21}): while the size of galaxies increases with stellar mass ($M_*$), ETGs show a steeper growth, but are on average smaller than LTGs at fixed $M_*$. This gap in size disappears at the high-mass end ($M\sim 10^{11}$~$M_{\odot}$) where the MSRs of the two populations are crossing \citep{van14, mow19b}. These differences strongly suggest that these two populations follow their own different evolutionary paths. Recent JWST observations from The Cosmic Evolution Early Release Science Survey (CEERS) show that star-forming and quiescent galaxies already have a different morphology and different Sersic index at $z=5.5$ \citep{war24}.\\

The size enhancement of ETGs observed with decreasing redshift may result from one or several mechanisms with different weighted contributions, but those processes are currently still debated. One of the most popular is the dry (i.e., gas-poor) major and minor merger scenario (e.g., \citealt{cox06, kho06, naa06a, bou07, naa09a, bez09, ose10, tru11, fer14, oog16}), where ETGs efficiently grow without rejuvenation of star formation. An analysis of the Millenium Simulation Data Base \citep{spr05, delu07} by \citet{oog13} reveals that major and minor mergers govern the evolution of ETGs, and several observations have shown that major mergers could indeed impact the size growth of massive early-type galaxies (e.g., \citealt{lin08, lop12, oog13, fer14}). Growth through minor mergers is, however, preferred since the growth in size is faster than the growth in mass \citep{hue13a}, and major mergers would lead to more massive galaxies than currently observed (e.g., \citealt{lop09}) and to a change in the MSR slope (e.g., \citealt{ber11, swe17}).\\
Another mechanism proposed for the size growth of massive spheroidal galaxies relies on an in situ process, where the quasi-adiabatic expansion is governed by the ejection of matter from the central to the outer parts of the galaxy due to AGN feedback, stellar winds, and supernovae explosions \citep{fan08, fan10, wil10, rag11}.\\
We previously mentioned that the mean size evolution does not automatically imply that ETGs grow through time. Indeed star-forming galaxies can experience quenching through tidal stripping (e.g., \citealt{lok20}), ram pressure stripping (e.g., \citealt{jos20}), or harassment (e.g., \citealt{bia15}). As a consequence, it is also possible that all these mechanisms listed above are of minor importance, and what drives the observed mean size growth in a population of ETGs is the incorporation of these recently quenched, and larger than quiescent, star-forming galaxies into the passive population. This effect, known as the progenitor bias, might be majorly responsible for the observed size growth of the passive population (e.g., \citealt{car13,bel15,gar16,gar17,gar19,mat22}).\\

How galaxies evolve is further complicated by the environment that may impact their evolutionary paths. Some properties, such as the star formation rate (SFR), color, and morphology are known to depend on the environment. The density-morphology relation \citep{dre80} shows that elliptical and spiral galaxies are preferentially located in high-density (HD) and low-density (LD) environments, respectively (e.g., \citealt{coo06, coo07, cap07, coi08, kov10a, mou18, pau19}). In addition, since the color and star formation activity of galaxies are correlated (e.g., \citealt{pog08,bai17}), an HD environment also hosts more red quiescent galaxies, while blue star-forming galaxies are found in LD environments. This indicates that ETG evolution may not be identical in different environments. We may expect ETGs in HD environments, such as those found inside groups and clusters, to undergo an accelerated size growth through more numerous major and minor merger events, or cluster-related processes compared to LD environments such as fields and voids (e.g., \citealt{cav92, mci08, fak09, dar10a, lin08, lin10, san13, delu14, yoo17}). At the same time, the merger rate in the highest density environments may be reduced due to the higher peculiar velocities of galaxies in these environments (e.g., \citealt{tra08,jia12,mat19})\\

As a consequence, observations of galaxies in different environments might reveal a difference in size growth, yet the works that have been done so far lead to divergent results. For instance, several studies found the impact of environment on ETGs to be negligible (e.g., \citealt{ret09,wei09,fer13,cap13,hue13a,hue13b,kel15,sar17,zan21}), some found ETGs to be larger in HD environments (e.g., \citealt{cim08,coo12,pap12,str13,bas13,lan13,ceb14,dela14,mei15,yoo17,cha18,hua18,wang20,noo21,afa23}), while other studies claim that ETGs in HD environment are smaller (e.g., \citealt{rai12, pog13, mat19}). It is, therefore, not entirely clear whether the environment has a measurable impact and how this impact, if any, translates in term of galaxy sizes. Differences could also arise from the methodology used to derive the stellar mass and the size, from the criterion adopted to select passive galaxies, or from the methodology used to quantify the environment.\\

The goals of the present paper are twofold. First, we study the MSR of quiescent galaxies at $0.5\le z \le 0.9$  from the final release of VIPERS \citep{guz13,gari14,sco18}, the largest existing spectroscopic survey in the redshift range $0.4\le z \le 1.2$, with a statistically significant mass-complete sample of spectroscopically measured galaxies. We then consider the impact of the environment on the quiescent population of VIPERS through the density contrast that was estimated for each galaxy \citep{cuc14,cuc17} and on the size growth in the redshift range $0.5\le z \le 0.9$.\\
The paper is organized as follows. In Section.~\ref{Sect:data} we describe VIPERS and the derivation of associated physical parameters as well as the selection of the mass-complete sample of quiescent galaxies. In Section.~\ref{Sect:analysis} we analyze the MSR of the quiescent population and the impact of the environment. In Section.~\ref{Sect:discussion} we discuss the results, and present the conclusions of this work in Section.~\ref{Sect:conclusions}.\\
Throughout this work, we use $\Omega_M=0.3$, $\Omega_{\Lambda}=0.7$, and $H_{0}=70$~km~s$^{-1}$~Mpc$^{-1}$. All magnitudes are given in the AB system \citep{oke74}.

\section{The VIPERS data}\label{Sect:data}
\subsection{The survey}
VIPERS \citep{guz13, gari14, sco18} is a spectroscopic survey based on the T005 release of the Canada-France-Hawaii Legacy Survey Wide (CFHTLS-Wide, \citealt{gor09}) catalog toward a subarea of the W1 ($\sim$16~deg$^{2}$) and W4 fields ($\sim$8~deg$^{2}$). To constrain the redshift range of target galaxies to $0.5\le z\le1.2$, a first selection was performed based on a limiting magnitude $i<22.5$~mag and on a \textit{gri} color-color selection. The half-light radius and reconstructed spectral energy distribution (SED) from CFHTLS \textit{ugriz} bands were used to perform preliminary stellar decontamination. These selection criteria were checked using two control samples from VVDS-Deep \citep{lef05} and VVDS-Wide \citep{gar08}. Removing stars and galaxies outside the redshift range of interest beforehand led to a much higher sampling rate ($\sim$47\%) compared to previous surveys.\\

Observations were performed with the VIsible Multi-Object Spectrograph (VIMOS, \citealt{lef03}), a four-channel imaging spectrograph with a 224~arcmin$^{2}$ field of view at the ESO Very Large Telescope (VLT). Using the multiobject-spectroscopic mode and the low-resolution red grism, the spectral coverage goes from 5500 to 9500~\AA\, in the observed frame with a resolution of $R=220$. Using the EZ code \citep{gar10,gari14}, a redshift ($z$), and a redshift flag ($z_{flag}$) indicating the degree of confidence in the redshift estimation and the agreement with the photometric redshift were assigned to each spectrum. Both quantities were then independently and carefully checked by two members of VIPERS.\\
In this work, we used the VIPERS PDR-2\footnote{\url{http://vipers.inaf.it/}} catalog consisting of 91~507 sources and the updated T007 version of the photometric catalog.

\subsection{Morphological parameters}

The morphological parameters of VIPERS galaxies were derived using the GALFIT code \citep{pen02} and are based on the CFHTLS-T006 images \citep{kry17}. A postage stamp centered around the galaxy was extracted from the CFHTLS tile with adequate size to estimate the background accurately. Using SExtractor \citep{ber96}, all objects except the galaxy were masked, and the estimated galaxy's parameters were used as a first guess for fitting a Sersic profile \citep{ser63} with GALFIT. The Point Spread Function (PSF) at the galaxy position used to convolve the Sersic profile is represented by a Moffat profile \citep{mof69} whose parameters were estimated from the isolated stars in each CFHTLS tile. The free parameters of the Sersic profile estimated by GALFIT are the Sersic index ($n$), the semi-major axis ($a_{e}$), the axial ratio ($b/a$), the effective radius ($R_{e}=a_e\sqrt{b/a}$) of the galaxy, and the continuum background. Based on the tests performed on the simulated galaxy images, the Sersic index $n$ and the effective radius $R_e$ are accurate to 33\% and 12\%, respectively, for 95\% of the VIPERS galaxies. More details can be found in \citet{kry17}.\\
We note that different definitions of the galaxy size are used in the literature, such as the effective radius (e.g., \citealt{dam22,dam23}), the semi-major axis (e.g., \citealt{van14}), a radius enclosing a certain light fraction (20, 50 and 80, 90\%, e.g., \citealt{ich12, mil19, mos20}) or sometimes using a mass fraction (e.g., \citealt{sue19,mil23}), which can lead to significantly different results concerning the size evolution of galaxies (e.g., \citealt{mil23}). In this work, we use the effective radius as the size of a galaxy.\\
Observations of galaxies at different rest-frame wavelengths also affect their measured size, as different wavelengths trace different galaxy components. The $i$-band observations, used to measure $R_e$, trace the (rest-frame) emission at 5000 and 3950~\AA\, at $z=0.5$ and $z=0.9$, respectively. To quantify the color gradient, we follow the procedure of \citet{van14} and compute the effective radius of galaxies as if they were observed at 5000~\AA, $R_{\mathrm{5000\AA}}$, considering $\Delta\textrm{log}\,R_{e}/\Delta\textrm{log}\,\lambda=-0.25$. The conversion factor between $R_e$ and $R_{\mathrm{5000\AA}}$ decreases with redshift from 1 down to 0.96 and shows no strong dependence on stellar mass. Overall, the correction is very weak for our sample of quiescent galaxies, being lower than 0.017~dex as the rest-frame $i$-band is close in wavelength to 5000~\AA\, for the redshift range studied in this work. We therefore do not expect the color gradient to significantly affect the conclusions of this work and do not perform any corrections.

\subsection{The spectral break $D_n4000$}

The spectral break $D_n4000$ is a discontinuity observed in the spectrum of galaxies around 4000~\AA\, due to the lack of emission from young and massive OB stars and the apparition of absorption lines from ionized metals in the stellar atmosphere of evolved stars. Consequently, $D_n4000$ has been widely used as an evolutionary tracer of the galaxies' stellar age (e.g., \citealt{bru83,bal99,kau03a,siu17,hai17,dam22,dam23}). The spectral break $D_n4000$ is defined as the ratio of the integrated spectrum over the wavelength ranges $4000$-$4100$~\AA\, and $3850$-$4100$~\AA\, from the narrower definition of \citet{bal99}, which is less impacted by reddening compare to the former definition of \citet{bru83}, although this should not be critical as quiescent galaxies are usually not very dusty. The corresponding uncertainty on $D_n4000$ is estimated by propagating the uncertainties of the spectrum over both wavelength ranges.\\
One downside of using $D_n4000$ as a proxy for the stellar population age is the existence of an age-metallicity degeneracy \citep{wor94}. For VIPERS data, \citet{siu17} studied the impact of different metallicities on the stellar age of quiescent galaxies and showed that it does not strongly affect their conclusions. We therefore expect that this degeneracy will also not affect our conclusions. When comparing samples of galaxies based on the stellar age, we however use samples whose $D_n4000$ distribution is non-contiguous to reduce this age-metallicity degeneracy (see Sect.~\ref{Sect:final_sample}).

\subsection{Physical properties of galaxies}

The $M_*$ and SFR used in this work were derived by \citet{mou16a} from SED fitting using LePHARE \citep{arn99,ilb06} with the galaxy evolution explorer (GALEX) far-ultraviolet (FUV) and near-ultraviolet (NUV), the CFHTLS $u$, $g$, $r$, $i$, $z$, the VISTA Deep Extragalactic Observations survey Z, Y, J, H and K, and the Wide-field InfraRed Camera K$_s$ bands from the photometric catalog of \citet{mou16b}. The fitting procedure was performed with an exponentially declining star formation history with 0.1~Gyr $<\tau<$ 30~Gyr, the \citet{bru03} stellar population library, two different metallicities ($Z_{\odot}$ and 0.4$Z_{\odot}$), three different extinction laws \citep{pre84,bou85,cal00,arn13} with $E(B-V)_{\textrm{max}}=0.5$, contributions from emission lines \citep{ilb09} and a \citet{cha03} initial mass function. More information about the SED fitting process can be found in \citet{ilb13} and \citet{mou16a}.

\subsection{Sample selection}

\subsubsection{Selection of a reliable sample}\label{subsubsect:selectionsample}

We first performed a selection on the redshift quality of VIPERS galaxies through their associated $z_{flag}$ (see \citealt{sco18} for a detailed explanation), keeping galaxies with $2\le z_{flag}\le 9.5$ (90 to 99\% confidence level). We then reduced the redshift range to $0.5\le z\le 0.9$, as the density estimation we used is complete up to $z=0.9$ \citep{cuc17}.\\
We kept galaxies with accurate morphological measurements ($0.2<n<10$, $R_e>0.5$~kpc and $b/a>0.1$, \citealt{kry17}). Small values of $n$ are unphysical and their corresponding normalization factor in the Sersic profile is less accurate \citep{cio99}. In addition, we imposed a signal-to-noise ($S/N$) threshold on $R_e$ of 3. We also required $S/N (D_n4000)>3$, as $D_n4000$ is used later as a secondary criterion for quiescent galaxies selection. Two galaxies having unphysical $R_e$ ($\ge 40$~kpc) were discarded, and galaxies with GALFIT convergence warning and outside the PSF mask region were excluded. In addition, galaxies for which the $i$-band measurement was higher than 22.5~mag were excluded.\footnote{All VIPERS primary targets have $17.5<i<22.5$ in the T005 release of CFHTLS, but their magnitudes may slightly differ in the T007 release that is used in this work} The reliable sample selected by the criteria described above contains 43~739 galaxies.\\

The sample of quiescent galaxies is primarily defined using the NUVrK diagram \citep{arn13} and the criterion established by \citet{mou16a}

\begin{equation}\label{eq:NUVrK_moutard}
\begin{aligned}
  (NUV-r)&>3.372-0.029 t_L\\
  (NUV-r)&>2.25(r-K_{s})+2.368-0.029 t_L, 
\end{aligned}
\end{equation}

where $t_L$ is the look-back time of each galaxy. Out of the 43~767 galaxies in the reliable sample, 6942 are found to be quiescent.

\subsubsection{Mass completeness}\label{subsect:mass_completeness}

To estimate the mass completeness limit in the VIPERS flux-limited sample ($i_{lim}=22.5$~mag), we applied the method of \citet{poz10} based on the mass-luminosity ratio ($M/L$) and on the stellar mass limit ($M_{lim}$). The latter is defined as the stellar mass a galaxy would have if its $i$-band measurement would be equal to the magnitude limit of the survey. Assuming a constant mass-to-luminosity ratio, (i.e., $M_*/L(i)=M_{lim}/L(i_{lim})$), 
$M_{lim}$ is given by
\begin{equation}\label{eq:mass_limit}
\textrm{log(}M_{lim})=\textrm{log(}M_{*})+0.4(i-i_{lim}).
\end{equation}

\begin{figure}
	\resizebox{\hsize}{!}{\includegraphics{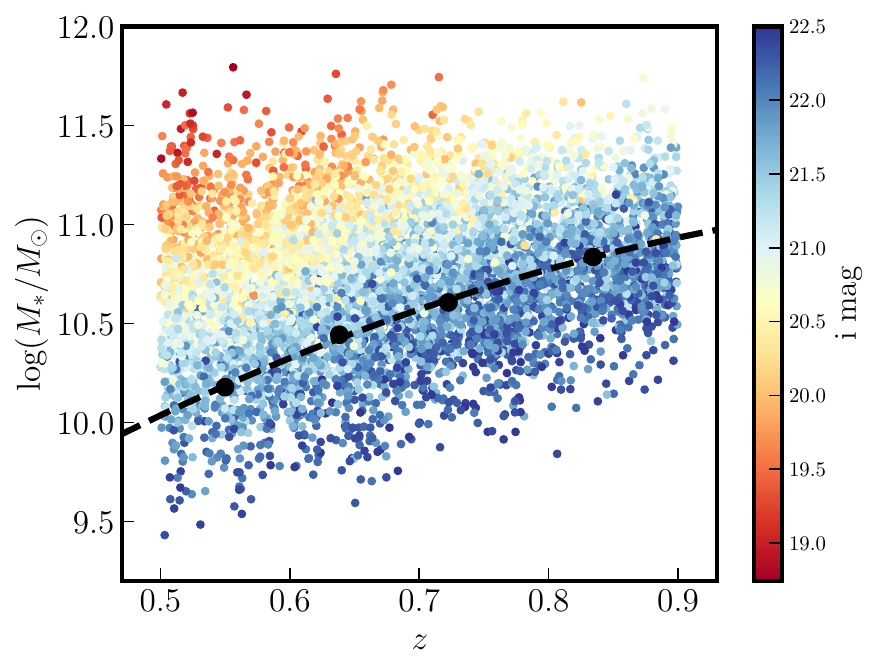}}
	\caption{Distribution of log($M_*/M_{\odot}$) with respect to $z$ where galaxies are color coded based on their $\textit{i}$-band magnitude. The black filled circles represent the completeness limit in each redshift bin and the dashed black line shows the completeness limit based on a 2$^{\textrm{nd}}$ order polynomial fit.}
	\label{fig:mass_completeness}
\end{figure}

We binned the sample per redshift quartile ($0.50\le z< 0.60$, $0.60\le z<0.68$, $0.68\le z< 0.78$, and $0.78\le z\le 0.90$) and selected the 20\% faintest galaxies in the $i$-band in each redshift bin. We rescaled their $M_{*}$ to $M_{lim}$ using Eq.~\ref{eq:mass_limit} and computed the $M_{*}$ completeness limit as the 90$^{th}$ percentile of the $M_{lim}$ distribution \citep{davi16}. A second-order polynomial fit was performed to obtain the completeness limit over the entire redshift range. The distribution of log($M_*/M_{\odot}$) with respect to $z$ and the mass completeness limit for quiescent galaxies are shown in Fig.~\ref{fig:mass_completeness}. The mass-complete quiescent sample contains 5124 quiescent galaxies (74\% of the reliable quiescent sample).\\
Our $M_{*}$ completeness limit at redshifts $z=0.60,0.72,0.84$ is $\textrm{log(}M_{lim}/M_{\odot}\textrm{)}=10.33$, $10.62$, and $10.84$, comparable to the estimation of \citet{davi16} with a difference of 0.06, 0.02 and 0.01~dex, respectively. These small differences can be attributed to a different estimation of $M_*$ (based on \textit{Hyperz} with different photometric bands for \citealt{davi16}), to the different NUVrK criterion used to select the quiescent galaxies, and to the selection criteria applied to obtain the reliable sample of galaxies.

\subsubsection{Final sample of quiescent galaxies}\label{Sect:final_sample}

\begin{figure}
	\resizebox{\hsize}{!}{\includegraphics{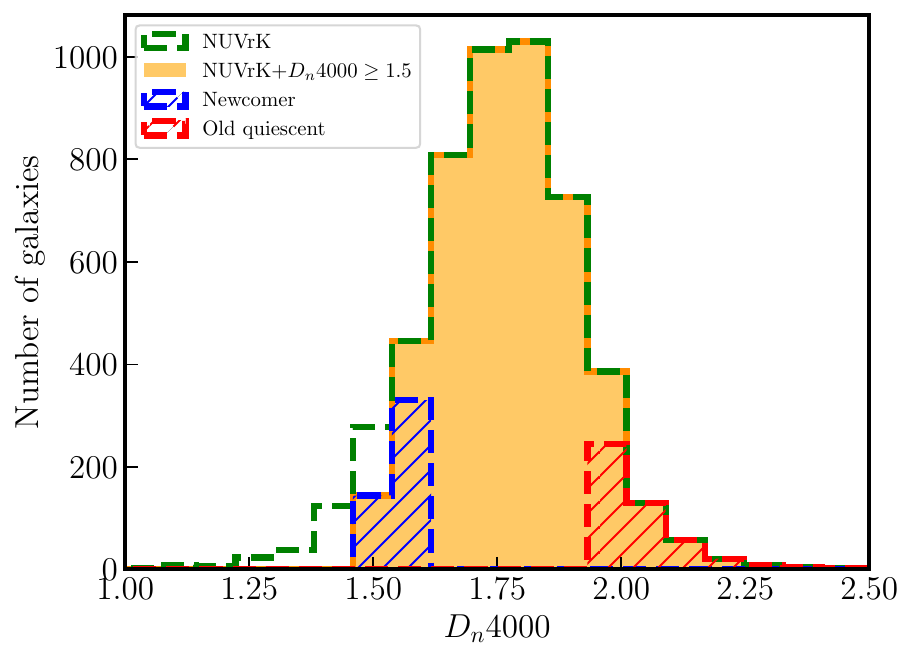}}
	\caption{$D_n4000$ distribution for the quiescent galaxies selected with the NUVrK diagram (green dashed), further constrained with $D_n4000\ge 1.5$ (yellow filled), newcomers (blue dashed), and old quiescent galaxies (red dashed).}
	\label{fig:histo_D4000_red}
\end{figure}

In addition to the selection based on the NUVrK diagram, as explained in Sect.~\ref{subsubsect:selectionsample}, the final mass-complete sample of quiescent galaxies was further constrained with $D_n4000$. Following \citet{dam23}, the sample of quiescent galaxies was restricted to $1.5<D_n4000<3$ (4786 galaxies). We also defined the recently quenched population as galaxies for which $1.5<D_n4000<1.6$ (hereafter newcomers, 474 galaxies). A total of 338 galaxies, quiescent following the NUVrK selection but star-forming following the $D_n4000$ one, were discarded. To perform a more robust comparison between the properties of newcomers and older galaxies, we also constructed a sample of old galaxies defined to have high $D_n4000$ and as much galaxies as the newcomer sample, leading to a sample of old quiescent galaxies with $D_n4000>1.95$. Comparing two non-contiguous samples with respect to $D_n4000$ will allow to weaken the dependence of $D_n4000$ on metallicity, the $D_n4000$ difference corresponding to an age difference rather than a metallicity difference, and this will strengthen the conclusions of such comparison. Figure~\ref{fig:histo_D4000_red} shows the $D_n4000$ distribution of the quiescent population selected with the NUVrK diagram before and after the $D_n4000$ constraint, and the newcomer and old quiescent populations.

\begin{figure*}
	\resizebox{0.5\hsize}{!}{\includegraphics{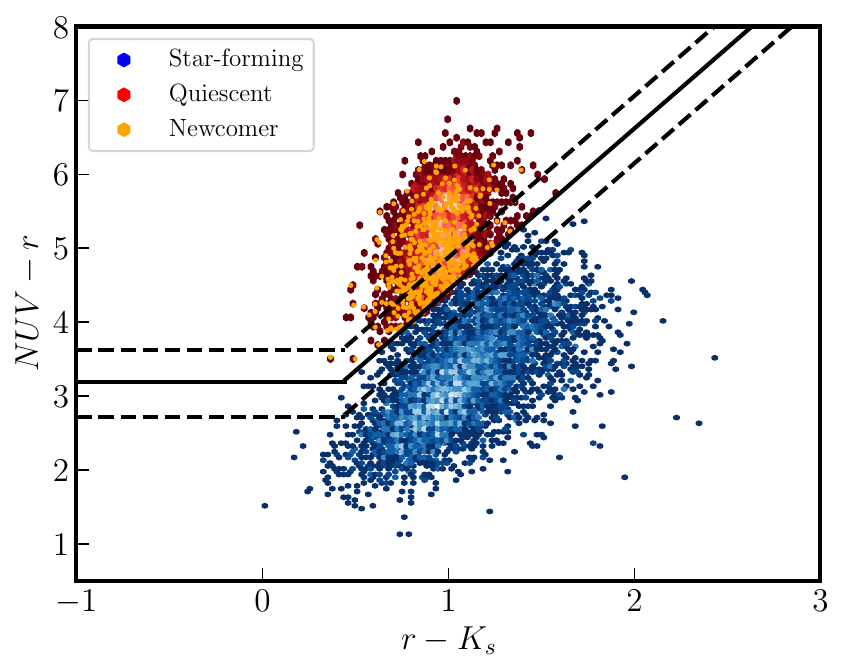}}\resizebox{0.5\hsize}{!}{\includegraphics{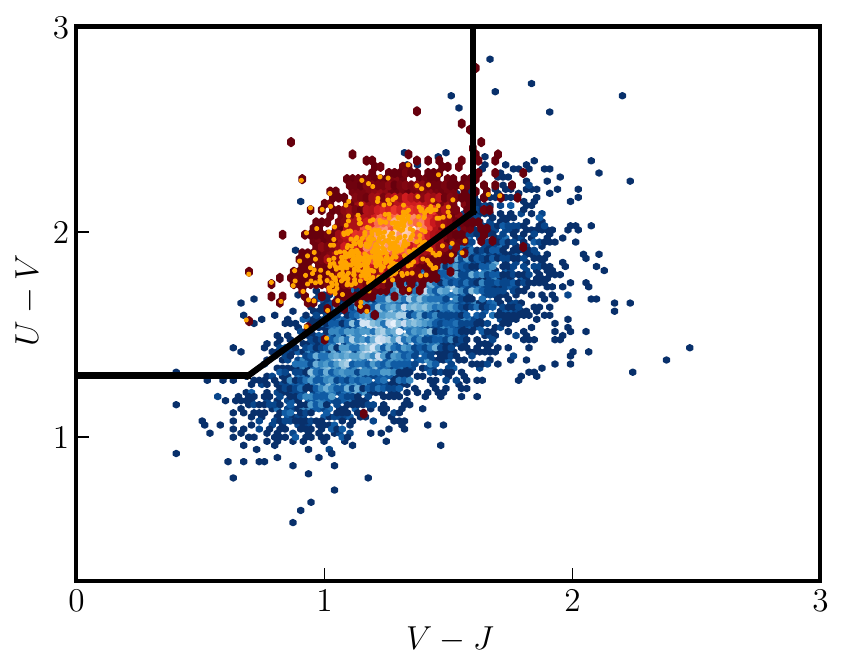}}
	\caption{NUVrK (left) and UVJ (right) diagrams for the star-forming, quiescent, and newcomer galaxies. The continuous black line represents the boundary between star-forming and quiescent galaxies (at $z\sim0.7$ for NUVrK). The green valley for the NUVrK diagram is enclosed by the black dashed lines.}
	\label{fig:NUVrK_UVJ}
\end{figure*}

In Fig.~\ref{fig:NUVrK_UVJ}, we show the star-forming, quiescent and newcomer galaxies on the NUVrK diagram with the thresholds from \citet{mou16a} at $z=0.7$. We also present the same three samples on the UVJ diagram \citep{wil09}, one of the most popular criterion used to disentangle the star-forming from the quiescent population, and note a good correlation between both selections, even if the separation is, as expected, not perfectly equivalent. We test the effect of different selections on the MSR in Sect.~\ref{Sect:impact_selectioncriteria}.

\subsection{Density estimation}\label{sect:density}

To measure the environmental impact on the quiescent galaxies, we used the density contrast $\delta$ estimated for VIPERS galaxies by \citet{cuc17}, and defined as
\begin{equation}\label{eq:density_contrast}
    \delta = \frac{\rho(r)-\left<\rho[r(z)]\right>}{\left<\rho[r(z)]\right>},
\end{equation}

where $\rho(r)$ is the local density at the comoving position $r$ and $\left<\rho[r(z)]\right>$ is the mean local density at redshift $z$.\\
The local density $\rho(r)$ of a galaxy was computed as the sum over all tracers (i.e., galaxies used to estimate the density) inside the smoothing filter divided by the selection function. The smoothing filter was chosen to be a cylindrical top-hat filter centered on the galaxy with a 1000~km~s$^{-1}$ half-length and a radius equal to the distance to the 5$^{\textrm{th}}$ neighbor ($D_{p,5}$). Using a cylindrical filter allows to reduce the effect of the peculiar velocities of galaxies and a varying radius allows to estimate the density on small-scales. Galaxy tracers are part of a volume-limited sample characterized by $M_B\leq -20.4-z$, ensuring completeness up to $z=0.9$ and a constant comoving number density with respect to the redshift. When estimating the local density, the selection function that includes all the different intrinsic selections of the VIPERS spectroscopic survey was not considered, but this was mitigated by adding to the volume-limited sample of tracers, all galaxies for which a photometric redshift was available. Local densities of galaxies for which part of the filter was out of the boundaries of the survey area are divided by the fraction of the cylinder found inside the survey. To compute $\left<\rho[r(z)]\right>$, the number density $n(z)$ was integrated over the cylinder height and divided by the survey area. We refer to Section~3 and Appendix.~A of \citet{cuc17} for more details about the density estimation, to \citet{cuc14} for a discussion about the inclusion of the selection function and to \citet{kov10b} for the estimation of $\left<\rho[r(z)]\right>$. In the following, we refer to $\delta$ simply as the density instead of density contrast. We note that this density contrast has already been used to study the quiescent and red nugget populations in VIPERS \citep{gar19,siu22,siu23,lis23}.\\

\begin{figure}
	\resizebox{\hsize}{!}{\includegraphics{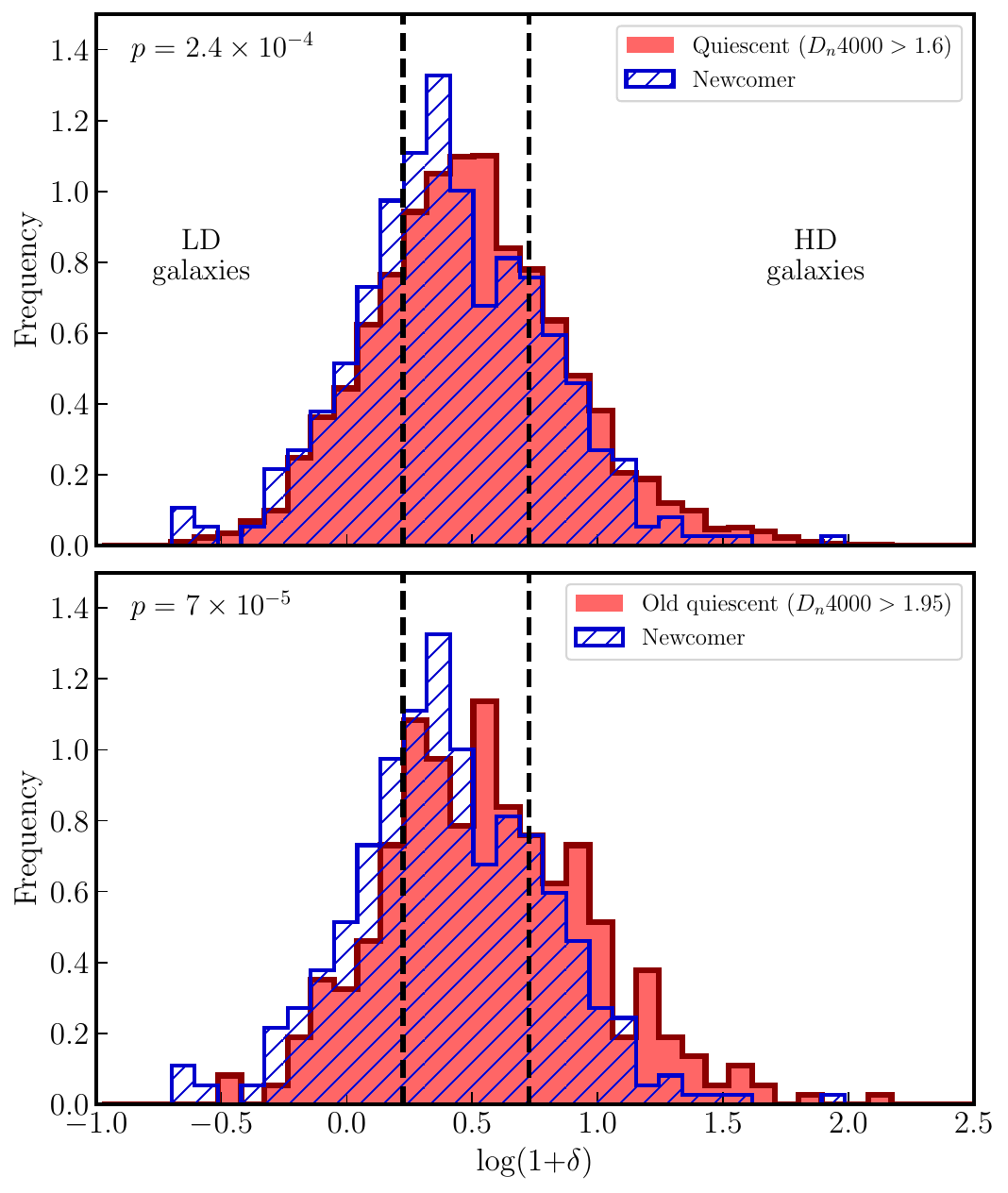}}
	\caption{Density distribution for quiescent (red) and newcomer (hash blue) galaxies. The black dashed lines indicate the LD and HD limits and the $p$-value from the KS test is indicated in top left. \textit{Top:} Histogram for newcomers and quiescent galaxies ($D_n4000>1.6$). \textit{Bottom:} Histogram for newcomers and old quiescent galaxies ($D_n4000>1.95$).}
	\label{fig:red_density_histogram}
\end{figure}

To study the impact of the environment, we selected a subsample of quiescent galaxies for which the filter covered at least 60\% of the field area. While boundary corrections were applied, we preferred to keep galaxies for which these corrections were low enough to ensure a good estimation of the density. For quiescent and newcomer galaxies, the samples are reduced to 3998 and 399 galaxies, respectively.\\
The thresholds used to separate LD and HD environments were defined as the first and third quartiles of the density distribution \citep{cuc17}. For the sample of quiescent galaxies used in this work, these thresholds are equal to log($1+\delta)=0.22$ and log($1+\delta)=0.72$ for the LD and HD environments, respectively, similar to \citet{cuc17} (0.23 and 0.73, respectively).
We show the histogram of the density for newcomer and quiescent galaxies with the density limits in Fig.~\ref{fig:red_density_histogram}. The newcomer and quiescent galaxies have a mean log($1+\delta)$ of 0.39 and 0.47, and a Kolmogorov-Smirnov (KS) test rules out that these two distributions come from the same parent population. Above log($1+\delta)=1.2$, the number of newcomers decreases abruptly, which is in agreement with the fact that these recently quenched star-forming galaxies should be less present in high-density environment where passive galaxies are preferentially found, following the morphology-density relation \citep{got03}. We reproduced the same analysis with the sample of old quiescent galaxies and found the same results: both populations do not have the same parent distribution and more older galaxies are found in the HD environment. The mean log($1+\delta)$ increases to 0.53, indicating that the older a quiescent galaxy is, the more likely it is to belong to a HD environment.


\begin{table}[h]
\caption{Number of galaxies in each sample}            
\label{tab:sample}      
\centering                          
\begin{tabular}{l|c}
Sample & Number of galaxies\\
\hline
\multicolumn{2}{c}{Initial samples (Sect.~\ref{subsubsect:selectionsample})} \\
\hline
VIPERS PDR-2 & 91~507 \\
Reliable  & 43~739 \\
Quiescent (NUVrK) & 6942 \\
\hline
\multicolumn{2}{c}{Mass-complete samples (Sects.~\ref{subsect:mass_completeness}, \ref{Sect:final_sample})} \\
\hline
Quiescent ($1.5\le D_n4000\le 3$) & 4786 \\
Newcomers ($1.5\le D_n4000\le 1.6$) & 474 \\
Old quiescent ($1.95\le D_n4000\le 3$) & 474 \\
\hline
\multicolumn{2}{c}{With reliable density estimation (Sect.~\ref{sect:density})} \\
\hline
Quiescent & 3998 \\
Newcomers & 399 \\
\end{tabular}
\end{table}

\section{Analysis}\label{Sect:analysis}

\subsection{MSR of the quiescent population}\label{Sect:MSR}

We analyze the MSR of quiescent galaxies in VIPERS, which is the spec-only sample that large at this redshift range. We divided the sample into $M_{*}$ quartiles ($\sim$ 1197 galaxies per bin) and estimated the median $M_{*}$ and $R_e$ in each bin. The uncertainties on $R_e$ are computed as the median absolute deviation divided by the square root of the number of galaxies in each bin. The general shape of the MSR is represented by a smoothly broken power law (e.g., \citealt{mow19b}) or a combination of two power law functions (e.g., \citealt{she03, fern13, lan15, fur17, ned21}), which for quiescent galaxies at $M_*>10^{10}M_{\odot}$ can be reduced to a single power law,

\begin{figure*}[t]
	\resizebox{\hsize}{!}{\includegraphics{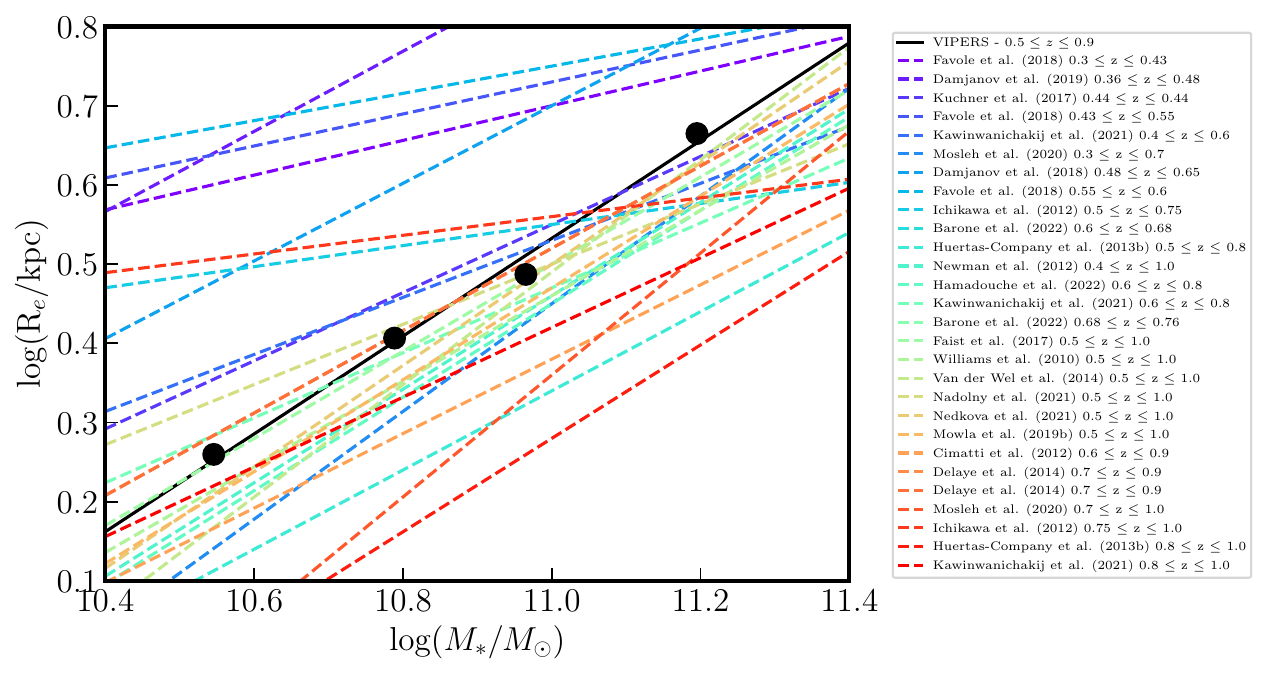}}
	\caption{MSR for quiescent galaxies from this work (black circles) and the literature (see Tab.~\ref{tab:MS_literature}) at $0.3\le z\le 1$ (blue to red dashed lines, sorted by redshift). The linear fit for the VIPERS quiescent sample is shown by the continuous black line.}
	\label{fig:mass_size}
\end{figure*}

\begin{equation}\label{eq:MR}
\textrm{log(}R_e\textrm{)}=\alpha\times \textrm{log}\left(\frac{M_{*}}{10^{11}M_{\odot}}\right)+\textrm{log(}A\textrm{)},
\end{equation}
where $\alpha$ is the slope and $\textrm{log}(A)$ is the intercept. We performed the fit using an orthogonal distance regression technique from the \textit{scipy} python package, taking into account the uncertainties on $R_e$. As noted by \citet{sar17}, an orthogonal distance or a linear least squares regression can lead to parameters that can be different by $\sim2\sigma$. We checked that the parameters for this MSR, and all MSRs that are derived in this work, were consistent within 1$\sigma$ when using both regression techniques. For the sample of quiescent galaxies, the fit gives a slope $\alpha=0.62\pm0.04$ and an intercept $\textrm{log(A}\textrm{)}=0.52\pm0.01$. The MSR for VIPERS along with several MSRs from the literature at $0.3\le z\le 1$ (see Tab.~\ref{tab:MS_literature} for a more extensive list) are shown in Fig.~\ref{fig:mass_size}, where we observe a good agreement with our work despite the different methodology used to select quiescent galaxies for instance. While the slopes of the different relations are roughly similar, except for the MSRs of \citet{ich12} and \citet{fav18}, we clearly observe the decrease in the intercept with redshift indicating that quiescent galaxies are smaller at higher redshift.\\

\begin{figure*}[ht!]
	\resizebox{\hsize}{!}{\includegraphics{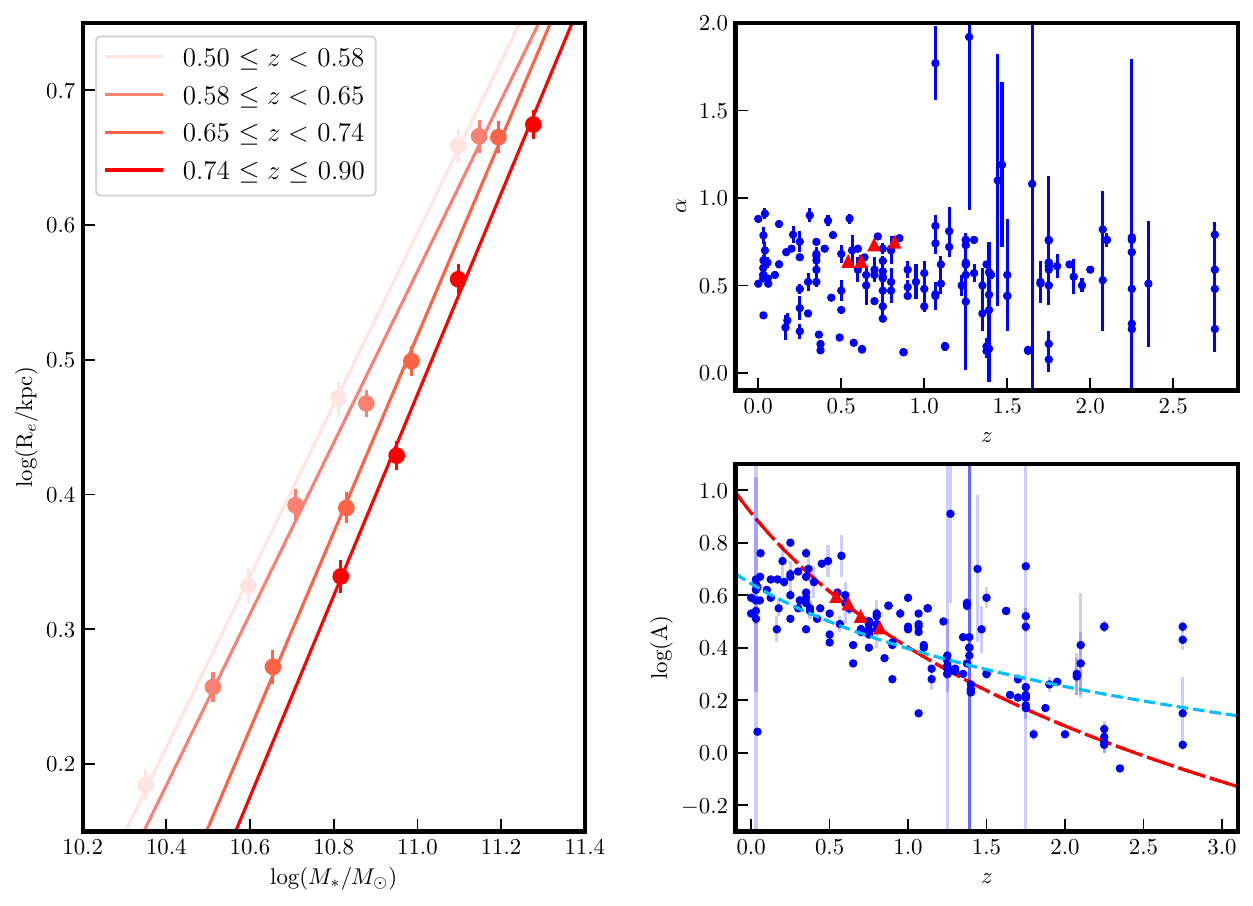}}
	\caption{MSRs for quiescent galaxies in different redshift bins (left) where the points represent the median values in each quartile and the solid lines represent the fits to the data bin. The uncertainty on the stellar mass, computed as the MAD divided by the square root of the number of galaxies, is of the order of $\sim 0.01$. Dependence of $\alpha$ and log(A) with redshift (right) for VIPERS (red triangles) and from the literature (blue circles). The fit for the intercept with redshift is shown as a red dot-dash (VIPERS) and blue dash line (all data points).}
	\label{fig:mass_size_z}
\end{figure*}

To check the evolution of the slope and the intercept with redshift, we divided the VIPERS sample into redshift quartiles and fitted the MSR using Eq.~\ref{eq:MR} in each redshift bin (the slopes and intercepts are given in Tab.~\ref{tab:MS_literature}). The MSRs for the four redshift ranges are shown in Fig.~\ref{fig:mass_size_z}~(left) where we observe that the intercept decreases with redsfhit between the first and fourth redshift bin ($\ge$3$\sigma$), while the slope appears to slightly increase. The redshift dependence of the slope and intercept for VIPERS are also shown on Fig.~\ref{fig:mass_size_z}~(right, red triangle), together with the slope and intercept values from the literature (Tab.~\ref{tab:MS_literature}). For a better comparison with the present study, a recalibration of the relations, affecting only the value of log(A), was performed, and already applied on Fig.~\ref{fig:mass_size_z} (bottom right). First, we corrected for the different $M_*$ normalization used in Eq.~\ref{eq:MR} which is equal to $10^{11}$~$M_{\odot}$ in this work. Secondly, we accounted for the fact that $R_e$ can be either defined as the semi-major axis size of the galaxy (e.g., \citealt{van14,bel15,fai17,mow19b,yan21,ham22}) or its circularized radius (e.g., \citealt{guo09,kro14,dam22,dam23}; this work), which differs by a factor of $\sim$1.4 \citep{dut11}. Therefore, the new intercept is given by $\alpha$log($10^{11}/N_M$)+log($A$)$-$log($N_R$), where $N_M$ is the mass normalization used in the original calibration and $N_R$ is equal to 1.4 if the original calibration uses the semi-major axis size and 1 if using the circularized radius. The uncertainty on the new intercept is estimated by keeping the relative uncertainty identical.\\

On Fig.~\ref{fig:mass_size_z}~(right), the slopes of VIPERS between the two first and two last redshift ranges seem to increase with redshift. However, when slopes from the literature are added, we observe that most of them are scattered between 0 and 1 (on average $0.56\pm 0.20$) without a clear trend, as shown by \citet{new12}, while the intercept is decreasing with redshift. This comparison suffers from different drawbacks such as different quiescent galaxies selection, color gradient \citep{van14}, stellar mass estimation (e.g., Cigale, EAZY, BAGPIPES), or radius estimation from a single/double Sersic profile. Correcting for these effects is not possible based on the fit of the MSR only. However, and despite several differences between different studies, we found the same trend for the evolution of the intercept with redshift as other works such as \citet{van14} or \citet{mow19b} where gradient color corrections were taken into account. Remarkably, such a trend is robust against all the different selection effects. \\
The evolution of the size at $M_*=10^{11}$~$M_{\odot}$ with redshift for the VIPERS sample is given by $R_e(z)=(8.20\pm 0.34)\times (1+z)^{-1.70\pm 0.08}$, in agreement with the evolution found in other works such as \citet{van14} and \citet{dam19} with an exponent of -1.48 and -1.6, respectively, and slightly higher than the fit from \citet{mow19b} with -1.25. We also fitted the evolution of log(A) with redshift using the data from the literature (including data from this work). For log(A) values without uncertainties, because they were not written or the MSR parameters were derived by eye, we set an arbitrary relative uncertainty of 50\%. Using data from the literature, the evolution is given by $R_e(z)=(4.76\pm 0.38)\times (1+z)^{-0.81\pm 0.12}$.\\
The mass-normalized radius ($R_e$ at $M_*=10^{11}$~$M_{\odot}$) of the population of quiescent galaxies increases by a factor of 1.3, from 3.01 to 3.91~kpc over 1.6~Gyr ($0.5\le z \le 0.9$), with a median size of 3.4~kpc at $z\sim 0.7$. These values are in agreement with other studies at the same redshift (e.g., \citealt{ham22}).

\subsection{The newcomer population}\label{sect:progenitor_bias}

As the redshift decreases, the passive population is constantly enriched with star-forming galaxies that have experienced a quenching process. Using $D_n4000$ together with the criterion from \citet[see also Sect.~\ref{Sect:final_sample}]{dam23}
allow us to study the properties of these newcomers at different cosmic times. Because star-forming galaxies are larger than the quiescent ones at fixed stellar mass, we would expect newcomers to be larger than the quiescent population. This effect is important to consider as the arrival of large quenched galaxies would enhance the mean size of the quiescent population without any size growth mechanisms needed, and could explain, at least partially, the apparent observed size evolution over cosmic time.\\

\begin{figure}
\resizebox{\hsize}{!}{\includegraphics{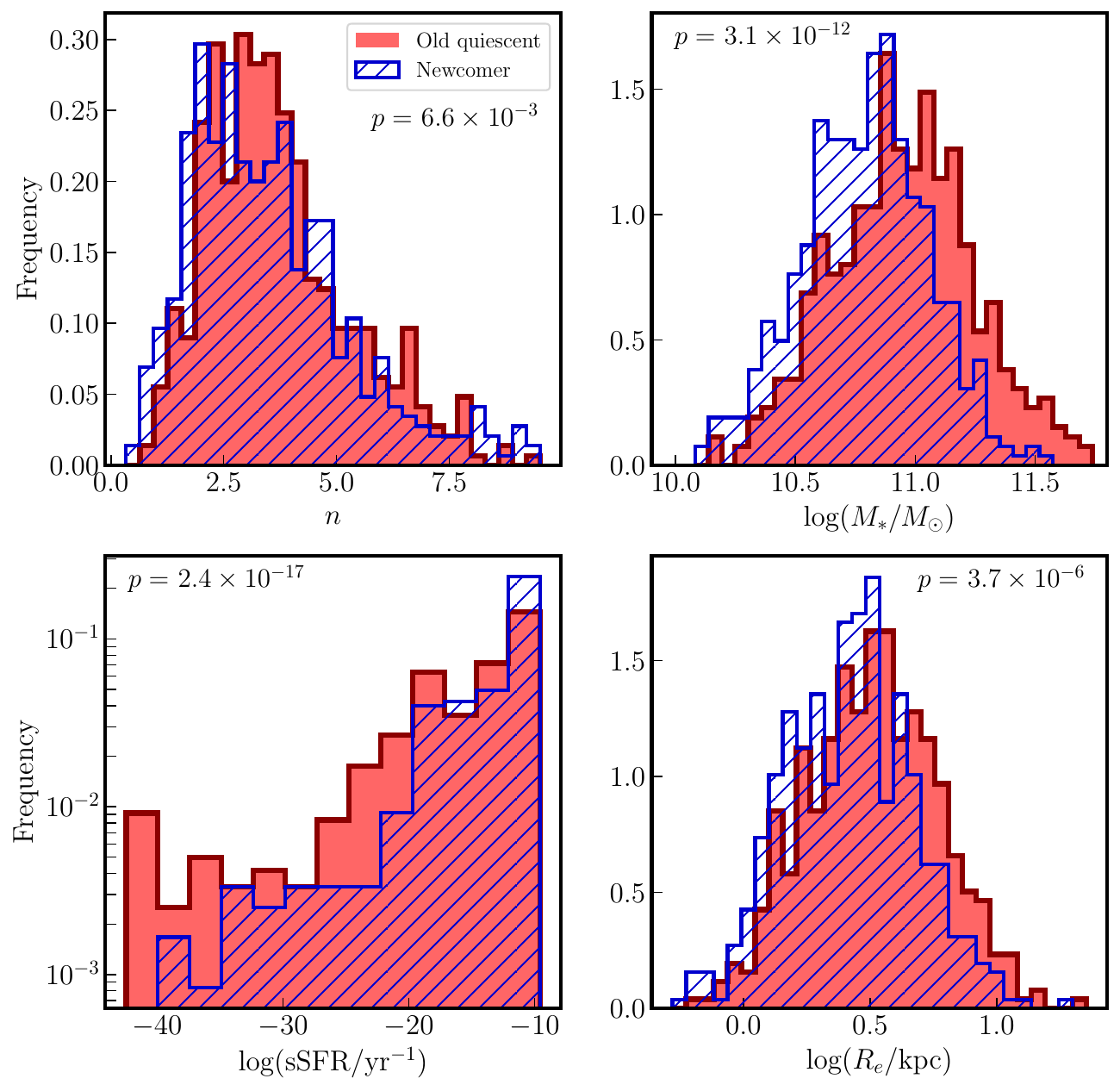}}
\caption{Distributions of $n$, $M_{*}$, sSFR, and $R_e$ for newcomer (hash blue) and old quiescent (red) galaxies.}
\label{fig:histo_red_new_trans}
\end{figure}

In the following, we compared the population of newcomers with the population of old quiescent galaxies defined in Sect.~\ref{Sect:final_sample}. The distributions of several properties shown in Fig.~\ref{fig:histo_red_new_trans} indicate that newcomers have a smaller Sersic index ($\overline{n_e}$ of 3.2 and 3.5), a smaller $M_*$ ($\overline{\textrm{log(}M_*)}$ of 10.8 and 10.9), and a larger star-formation activity ($\overline{sSFR}$ of -11.9 and -13.3), in agreement with being star-forming galaxies that have recently quenched. The log($R_e$) distribution is shifted toward lower size for newcomers, while the mean size of star-forming galaxies is higher than the quiescent ones. Therefore, the quenching process of these star-forming galaxies appears to go along with a size diminution, such as disk fading for instance. A KS test indicates that both populations, either based on their $n$, $M_*$, $R_e$ and sSFR are drawn from different samples.

If progenitor bias is responsible for the average size evolution of the quiescent population, the average size of newcomers should be larger compared to that of old quiescent galaxies at fixed mass. We constructed mass-match (MM) samples (see Sect.~\ref{sect:environment}) of the newcomer and old quiescent populations and show their size distributions in Fig.~\ref{eq:density_contrast}. The size distribution of newcomers is similar to that of the old quiescent galaxies and this is also confirmed by the KS test indicating that these distributions may have the same origin. It is therefore not clear if the progenitor bias could be the dominant process of size evolution, contrary to what is observed in \citet{car13,gar17,gar19} and \citet{dam19} as it would require the mean size of the newly quenched galaxies to be larger in order to explain the size increase of the whole quiescent population. We also checked the size distributions in different redshift bins ($0.50\le z \le 0.62$, $0.61\le z \le 0.68$, $0.68\le z \le 0.76$ and $0.76\le z \le 0.90$) and found the same trend: the size distribution is similar with $p=0.05,0.5,0.9$ and 0.4 in each redshift bin.\\

\begin{figure}
\resizebox{\hsize}{!}{\includegraphics{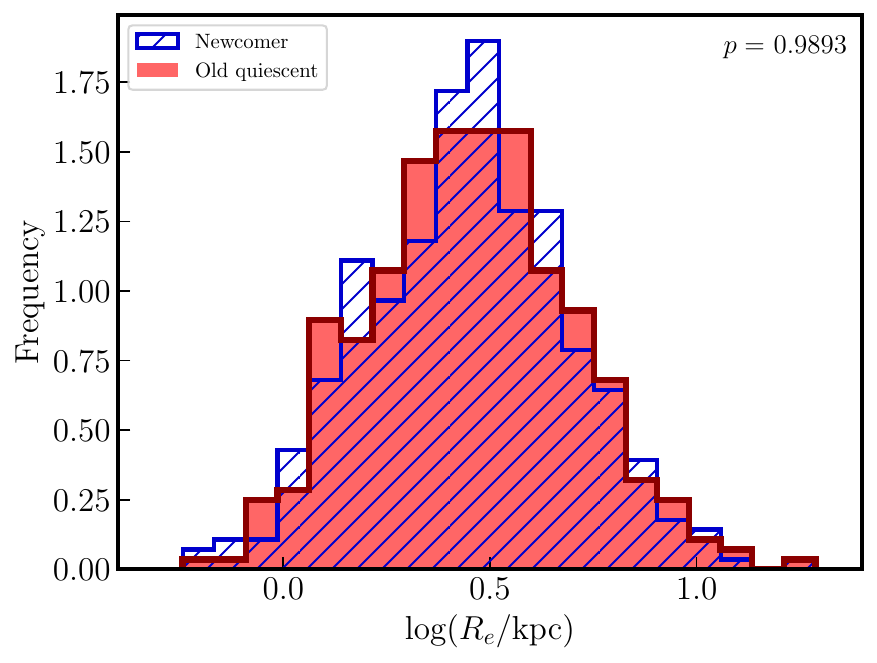}}
\caption{$R_e$ distribution for old quiescent (red) and newcomer (hash blue) galaxies.}
\label{fig:histo_red_new_size}
\end{figure}

\subsection{Impact of the environment}\label{sect:environment}

We made use of the density estimations computed by \citet{cuc14,cuc17} and presented in Sect.~\ref{sect:density} to study the environmental impact on the mean properties of the quiescent population. We recall that the sample of quiescent galaxies with accurate density estimations is reduced due to the removal of those having more than 40\% of their filter outside the surveyed area. Additionally, the samples used in the following analysis were further reduced as the HD and LD samples were defined as galaxies below and above the first and third quartile of the density distribution, respectively (see Fig.~\ref{fig:red_density_histogram}).\\

\begin{figure}
    \resizebox{\hsize}{!}{\includegraphics{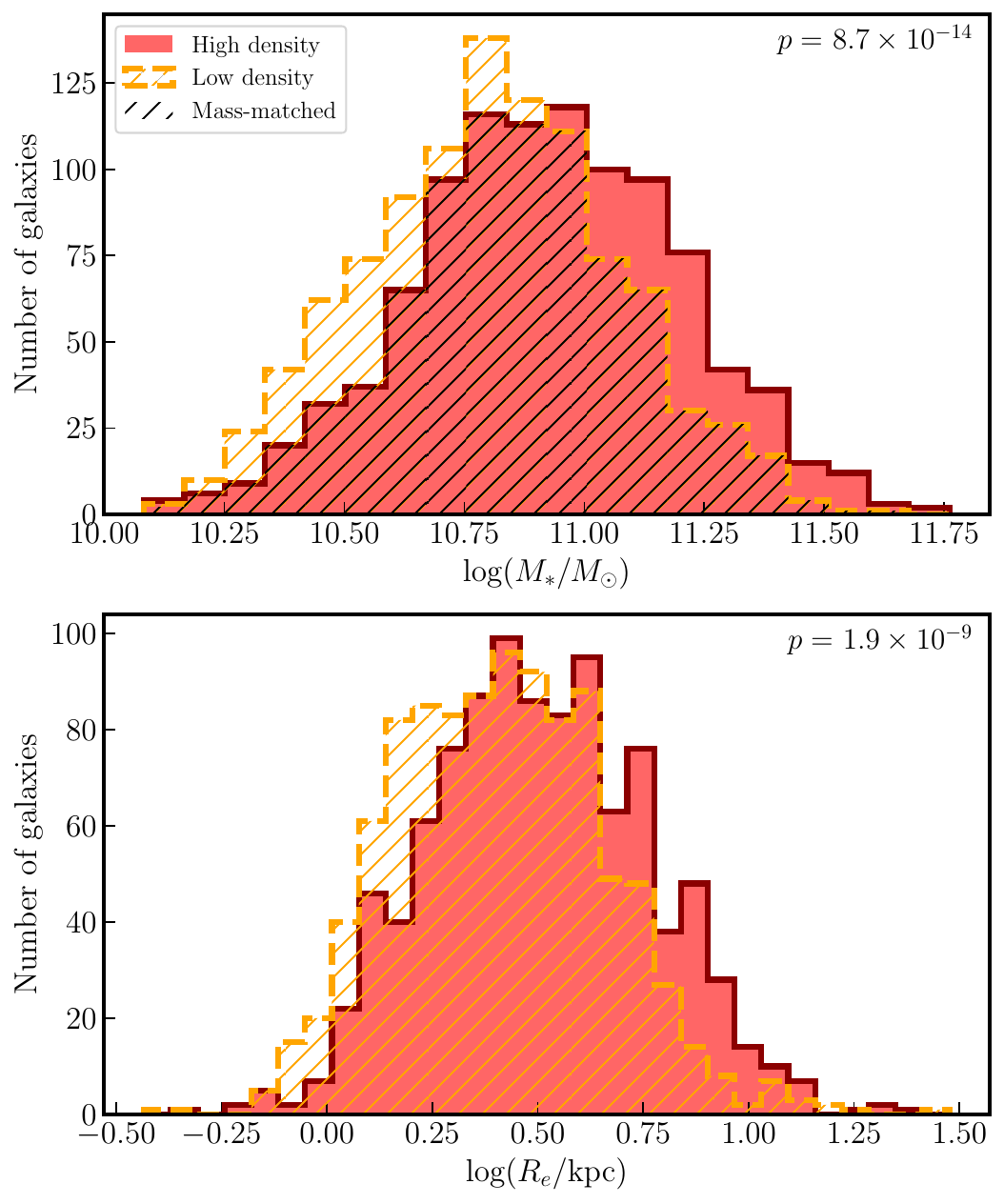}}
	\caption{Distributions of log($M_*$) for the HD (red), LD (dashed orange), and MM (hashed black) samples, and distribution of log($R_e$) for the HD (red) and LD (dashed orange) samples.}
	\label{fig:histo_Mass_Re_dens}
\end{figure}

We show the $\textrm{log(}M_*)$ and $\textrm{log(}R_e)$ distributions for the sample of quiescent galaxies in the HD and LD environments in Fig.~\ref{fig:histo_Mass_Re_dens}. Galaxies in the HD environment are found to be more massive and larger ($\overline{\textrm{log(}M_*)}=10.9$ and $\overline{\textrm{log(}R_e)}=0.5$) while low-mass and small galaxies dominate the LD environment ($\overline{\textrm{log(}M_*)}=10.8$ and $\overline{\textrm{log(}R_e)}=0.4$). Above $\textrm{log(}M_*)\sim 10.9$ and $\textrm{log(R}_e)\sim 0.4$, more than 50\% of the galaxies in each bin are found in HD environment and this fraction increases to 75\% at $\textrm{log(}M_*)\sim 11.4$ and $\textrm{log(R}_e)\sim 1.4$. Quantitatively, these differences in $M_*$ and $R_e$ for both environments are seen from the $p$-value of the KS test indicating that the $\textrm{log(}M_*)$ and $\textrm{log(}R_e)$ distributions in HD and LD environments are different. The difference in $M_*$ and $R_e$ between HD and LD environments is further observed from a two-dimensional two-sample KS test \citep{pea83} giving $p=5.1\times 10^{-11}$.\\
Galaxies with larger mass and size in clusters than in fields have been observed in \citet{rai11} and \citet{sar17} at $z\sim 1.3$. We do not observe a sharp cut in the $\textrm{log(}M_*)$ and $\textrm{log(}R_e)$ distributions for low-mass and smaller galaxies (large and massive) galaxies in HD (LD) environment, indicating that those exist but are more rare. The formation of massive galaxies seem to be favored in the highest-density environment, but to decipher the true impact of environment, we have to ensure that the observed difference, if any, is exclusively due to the environment.

The evolution of quiescent galaxies can be impacted by in situ processes depending on the mass of the galaxies or by environmental processes. To ensure that any changes observed between the LD and HD environments would be solely due to the environment, we created two new samples that we refer to as MM HD and LD samples. By definition, these MM samples have the same stellar mass distribution (Fig.~\ref{fig:histo_Mass_Re_dens} (top), the $M_*$ distribution of the MM samples is the intersection between the LD and HD histograms) and allow us to study the average size of the galaxies in both environments independently of their $M_*$. To create these MM samples, we divided the HD and LD samples into 20 equal-width $M_*$ bins and populated them so that the same $M_*$ bin in HD or LD environment has the same number of galaxies. Because the LD and HD samples do not have the same $M_*$ distribution, it follows that only a fraction of LD galaxies will end up in the LD MM sample in the low-mass regime and the same for HD galaxies in the high-mass regime (see Fig.~\ref{fig:histo_Mass_Re_dens} top). To take into account this random sampling we created 500 MM samples, both for the HD and LD environments. The physical properties were derived as the mean over the 500 MM samples and the uncertainty was estimated as the quadratic sum of the uncertainty due to the scatter in the size distribution and the uncertainty due to the random sampling.\\

\begin{figure}
    \resizebox{\hsize}{!}{\includegraphics{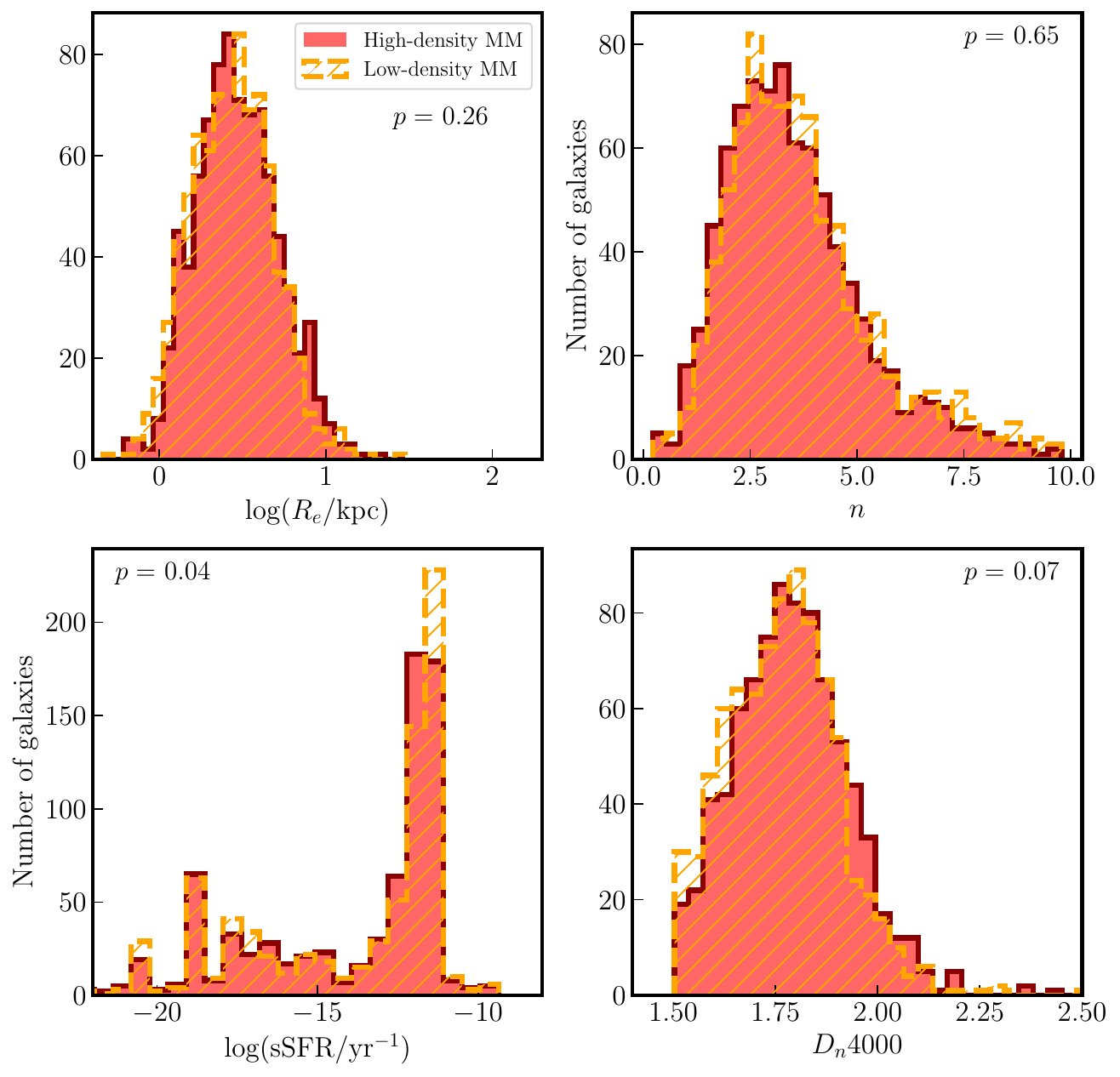}}
	\caption{Distributions of log($R_e$), $n$, log(sSFR), and $D_n4000$ for the HD (red) and LD (orange hash) MM samples. The average $p$-value of the KS two-sample test between the HD and LD galaxies for the 500 random samplings is shown at the top of each plot.}
	\label{fig:histo_red_dens}
\end{figure}

We show in Fig.~\ref{fig:histo_red_dens} the average distributions of $R_e$, $n$, sSFR, and $D_n4000$ in HD and LD environments over the 500 random samplings with their associated $p$-value. The $p$-value associated with the distribution of $R_e$, $n$, and $D_n4000$ of galaxies in LD and HD environments show that we cannot reject the hypothesis that they come from the same distribution. However, the $p$-value of the sSFR distribution weakly suggests that this property differs in the LD and HD samples, with galaxies in LD environment having a higher sSFR while remaining quiescent since the peak of the distribution is lower than -11 \citep{sal18,fig22,pea23}. The highest sSFR observed in LD could arise from the predominance of newcomers having a higher sSFR (see Figs.~\ref{fig:red_density_histogram}, \ref{fig:histo_red_new_trans}) and not specifically to processes related to the environment. We created LD and HD MM samples for quiescent galaxies with $D_n4000>1.6$ and $D_n4000>1.95$, as in Fig.~\ref{fig:red_density_histogram}, and recomputed the $p$-value for the sSFR and $D_n4000$ distributions. The $p$-values are equal to 0.06 and to 0.75 suggesting that the highest sSFR and lower $D_n4000$ is not a direct consequence of the environment, but due to the predominance of newcomers in the LD environment.\\

\begin{figure}
	\resizebox{\hsize}{!}{\includegraphics{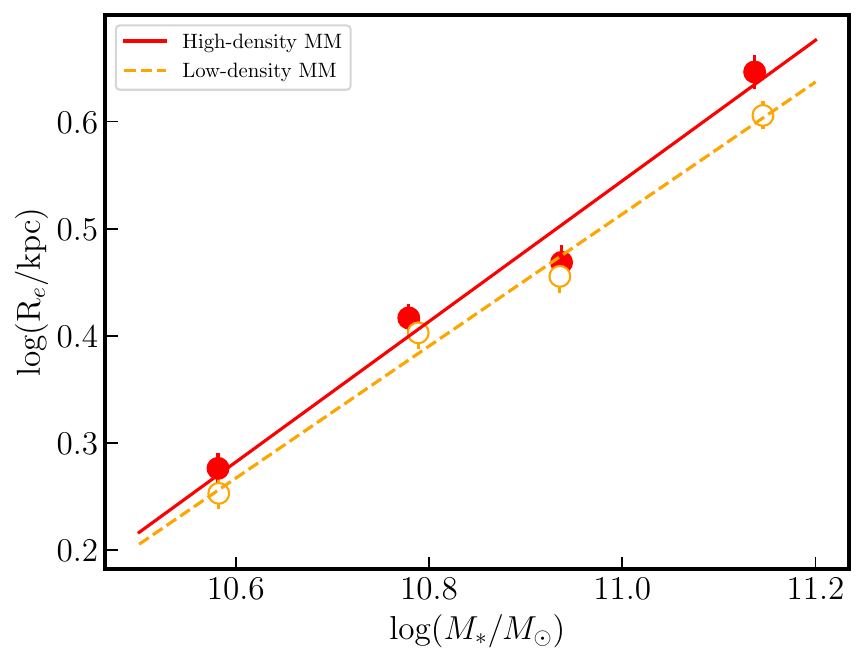}}
	\caption{MSR for HD and LD quiescent galaxies in the MM samples. Red filled and orange circles represent the median $M_{*}$, and the associated linear fits are shown as a red line and orange dashed line.}
	\label{fig:MS_MM}
\end{figure}

\begin{figure*}
	\resizebox{\hsize}{!}{\includegraphics{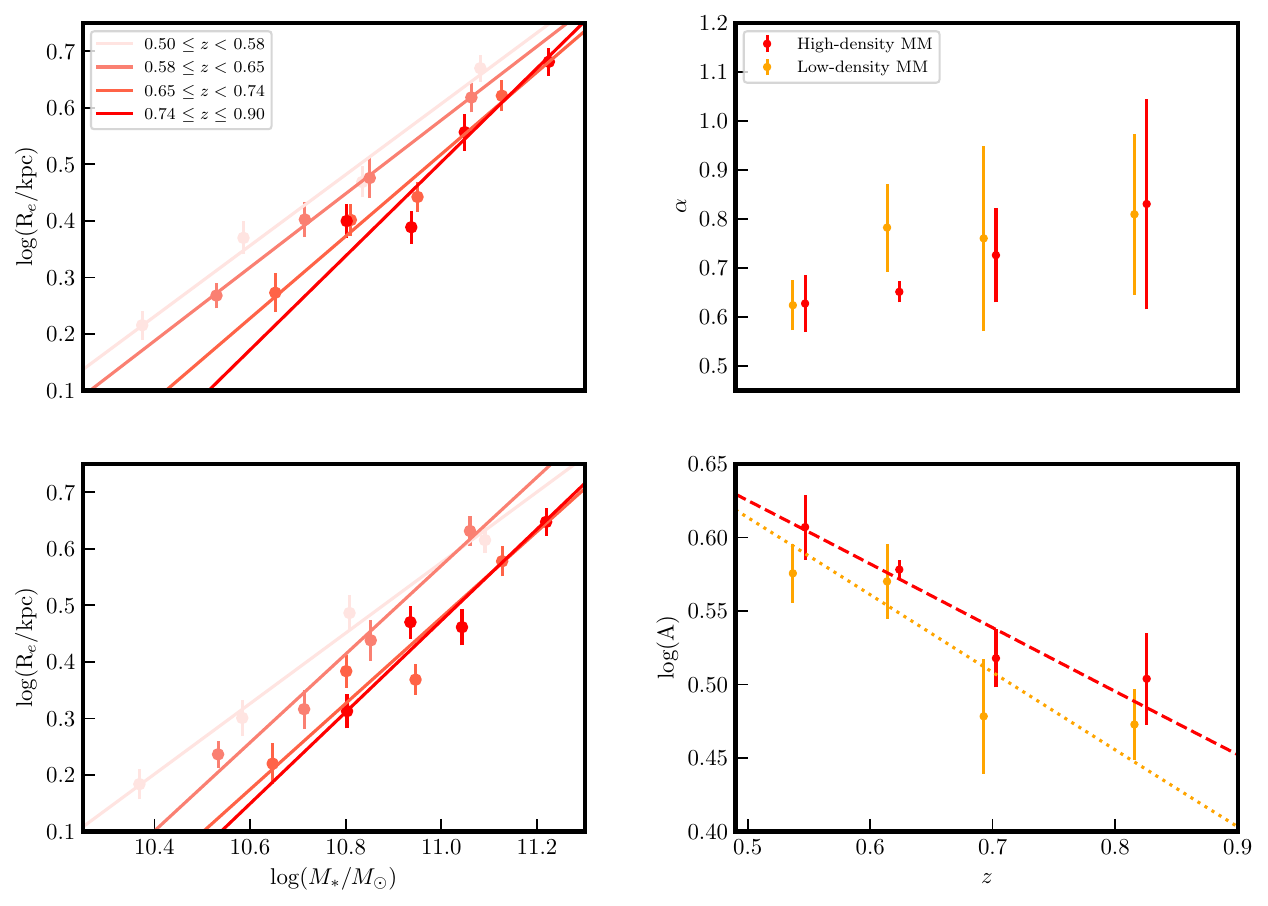}}
	\caption{MSR for quiescent galaxies in different redshift bins for the HD (top left) and LD (bottom left) environments, and $\alpha$ (top right) and log(A) (bottom right) as a function of redshift for the HD (red) and LD (orange) environments. The fitted evolutionary trends are shown as red dashed and orange dotted lines. The data showing the evolution of $\alpha$ and log(A) between LD and HD are shown with a slight offset to make the errorbars visible.}
	\label{fig:MShdld_MM}
\end{figure*}

The MSRs for the MM HD and LD samples are shown in Fig.~\ref{fig:MS_MM}, and the parameters of the fits are given in Tab.~\ref{tab:MS_literature}. At $M_*\le 10^{11}$~$M_{\odot}$, the sizes are very similar in the HD and LD environments within $1\sigma$ and we observe a larger difference of size at $M_*\ge 10^{11}$~$M_{\odot}$ although they agree within $2\sigma$. Therefore, the impact of environment-dependent mechanisms increasing the size of ETGs (such as mergers) does not seem to be efficient enough at high density to cause a significant size difference between both environments.\\
On Figure ~\ref{fig:MShdld_MM} we show the MSRs in four redshift bins (as in Fig.~\ref{fig:mass_size_z}) in the HD and LD environments, as well as the evolution of the slope and the intercept. We observe in both environments the same trend as shown in Fig.~\ref{fig:mass_size_z} (right): the constant slope and the intercept decrease with redshift. In addition, this trend is found to be independent of the environment.

\subsection{The $D_n4000$-$M_{*}$ relation}

\begin{figure*}[!ht]
\resizebox{0.5\hsize}{!}{\includegraphics{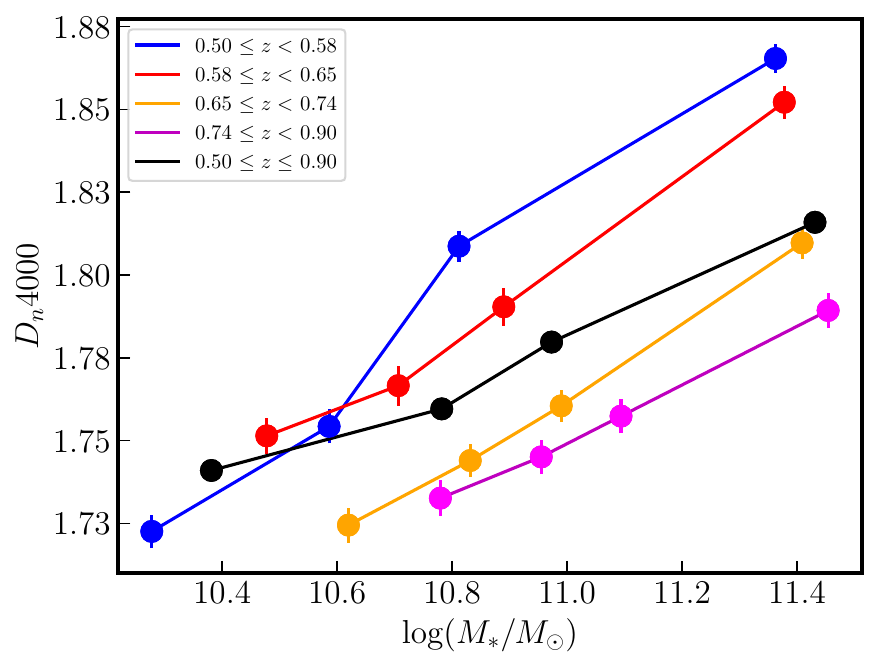}}\resizebox{0.495\hsize}{!}{\includegraphics{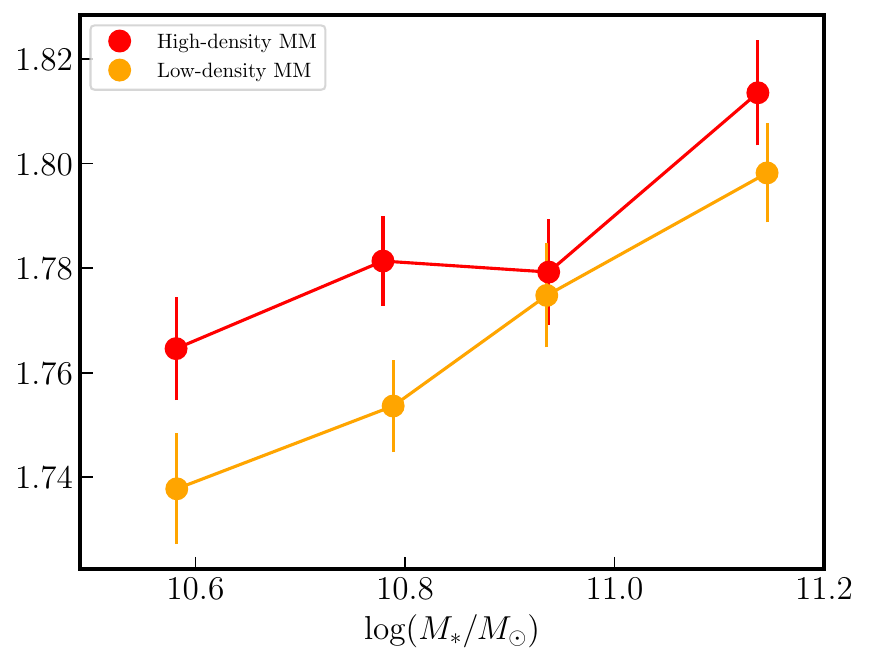}}
\caption{Relation between $D_n4000$ and log($M_*/M_{\odot}$) for the quiescent population in several redshift ranges (left), and for the HD (red) and LD (orange) environments (right).}
\label{fig:ReM_D4000_dens}
\end{figure*}

The $D_n4000$ spectral index has been shown to increase with stellar mass and to be offset with redshift: high-mass galaxies are older than the low-mass ones and galaxies at fixed stellar mass are older at low redshift \citep{kau03a, bri04, hai17, siu17, wu18a}. In VIPERS, this relation has been briefly studied by \citet{hai17} and more in-depth by \citet{siu17} for a slightly lower number of quiescent galaxies using stacked spectra in the redshift range $0.4\le z\le 1.0$. They found a positive slope independent of redshift and the $D_n4000$ increase with $M_*$ is explained by the aging of the stellar population with a non-negligible contamination from metallicity. However, the observed shift of the relation with redshift does not seem to be fully consistent with a simple passive evolution. By estimating the formation time, $z_{form}$, as the time of the latest burst of star formation, \citet{siu17} found that massive galaxies formed around $z_{form}=1.7$ while the low-mass ones formed at $z_{form}=1.0$. The fact that $z_{form}$ increases with $M_{*}$ is a signature of downsizing.\\
In this work, we found the same trends with $M_*$ (Fig.~\ref{fig:ReM_D4000_dens} left) with slightly higher $D_n4000$ values, despite the different selection criteria used to select the quiescent population. Contrary to \citet{siu17}, we find a weak increase in the slope with decreasing redshift, ranging from $0.086\pm 0.006$ to $0.13\pm 0.02$ and equal to $0.074\pm 0.006$ for the entire redshift range. This evolution of the slope remains in agreement with the SDSS slope of 0.15 \citep{hai17} or $0.141\pm 0.002$ \citep{siu17}. This evolution could be partially due to a progenitor bias with newly quenched galaxies entering the passive population at low $M_*$ and low $D_n4000$, steepening the slope as the redshift decreases.\\

To see how the environment impacts the $D_n4000-M_*$ relation, we used the MM samples that were defined in Sect.~\ref{sect:environment} and computed the mean $D_n4000$ in each bin of mass over the 500 MM samples. In Fig.~\ref{fig:ReM_D4000_dens} (right), we show the $D_n4000-M_*$ relation for both environments. In the HD and LD environments, $D_n4000$ increases with stellar mass, as for the whole quiescent population. We do not observe a significant difference for the age of galaxies at fixed $M_{*}$ between both environments. If the average size evolution was mainly due to mergers, we would have expected the age of the galaxies in HD to be higher than in LD while if the size evolution was mainly due to the progenitor bias, the age offset would have been smaller due to the addition of newly quenched star-forming galaxies. This would be in agreement with \citet{gar19} who found that the size evolution is mainly governed by progenitor bias rather than environmentally-dependent processes such as mergers at log($M_*/M_{\odot}\ge 11$). As in Sects.~\ref{sect:environment} and \ref{sect:density}, we constructed the $D_n4000-M_*$ relation in LD and HD environments with the samples of quiescent ($D_n4000>1.6$) and old quiescent galaxies ($D_n4000>1.95$) to check the impact of newcomers. In this case, no offset is observed and the largest age difference at low mass in Fig.~\ref{fig:ReM_D4000_dens} between the LD and HD environments likely originates from the galaxies that recently quenched to join the quiescent population. This is in agreement with the fact that newcomers are more present in the LD environment, which lower the $D_n4000$.

\section{Discussion}\label{Sect:discussion}

\subsection{How the quiescent selection criterion impacts the MSR}\label{Sect:impact_selectioncriteria}

How the quiescent population is selected is crucial as we want to simultaneously obtain the largest sample of quiescent galaxies with the lowest contamination by blue star-forming, dusty star-forming galaxies, and active galactic nuclei (AGNs). Several methods are available in the literature to perform such separation. While a morphological classification between ellipticals and spirals can be done at low redshift, classifications at higher redshift are often based on the observed bimodal distribution of physical parameters (sSFR, colors, Sersic index, $D_n4000$). As expected, applying different selection methods on the same sample leads to different samples of quiescent galaxies. In addition, the quiescent population resulting from one selection is also dependent on the procedure used and assumptions made to estimate the required physical parameters, such as using different SED fitting routines, different dust attenuation laws, or different SFR tracers.\\

The bimodality of the $D_n4000$ distribution has been used to disentangle the quiescent and star-forming populations with old and young stellar populations, respectively (e.g., \citealt{dam19,dam22,dam23,ham22}, this work). The evolution of $D_n4000$ in VIPERS has been investigated by \citet{hai17}, who found that the cut from \citet{kau03a} at $D_n4000=1.55$, was in good agreement with the observed bimodal distribution. The morphology-color relation can also be used to disentangle the two populations as elliptical and spiral galaxies are also well defined from their Sersic index \citep{tru01a,mcl13,del14}. \citet{kry17} found the mean $n$ for disk-like and spheroid-like galaxies in VIPERS to be between 0.81-1.11 and 2.42-3.69, respectively. Star formation tracers such as [O{~\sc{ii}}] or the sSFR can also be used as a way to separate both populations (e.g., \citealt{fra07,mig09,gar14,zan16,sal18,pea23}).\\

The $U-V$ color bimodal distribution associated with a constant color cut has been used to select quiescent galaxies (e.g., \citealt{bel04b}), but the lack of redshift evolution in this cut makes this selection too restrictive, excluding less luminous and less massive quiescent galaxies. This $U-V$ bimodal distribution with a redshift-dependent color cut has been applied to VIPERS in \citet{fri14}.\\
Criteria based on optical colors only (e.g., $U-V$, $U-B$, UVB diagram) are impacted by dust absorption, leading to higher contamination of the quiescent sample from dusty star-forming galaxies. To account for this effect, diagnostics including infrared rest-frame bands were introduced to allow for a more accurate selection, such as the UVJ (e.g., \citealt{wil09,van14,mow19b,ham22}), NUVrJ (e.g., \citealt{ilb10}) or NUVrK (e.g., \citealt{arn13,fri14,davi16,mou16a}) diagrams. Other color selections, that were perhaps less employed, include the BRI \citep{kuc17}, urz \citep{kaw21}, the $U-B$ \citep{pen10} or NUVr \citep{hue13b} diagram. The UBJ diagram for VIPERS galaxies shows a bimodal distribution but was not used directly for selection of quiescent/star-forming galaxies \citep{kry17}. 
VIPERS galaxies were also classified by \citet{siu18} using a Fisher Expectation-Maximization unsupervised machine learning algorithm into 12 classes sharing similar physical and spectral properties. While this kind of technique is difficult to apply, in particular to relatively small samples of galaxies, we decided to add this selection as clustering methods will be more and more applied for future large surveys. It is therefore interesting to see the robustness of such selections.\\

In the following, we compared the level of agreement between different star-forming/quiescent galaxies selections and examined the impact of these different selections on the MSR. To this end, we used the NUVrK boundary between quiescent and blue galaxies given by Eq.~\ref{eq:NUVrK_moutard} \citep{mou16a}, the $D_n4000$ criterion from \citep{dam23}, the UVJ criterion from \citet{ham22}, the redshift dependent $U-V$ color cut from \citet{fri14}, the $n$ limit from \citet{kry17}, the NUVrJ selection from \citet{ilb10}, and the clustering method from \citet{siu18}. Similarly to \citet{fri14}, we defined the completeness as the fraction of quiescent galaxies according to the main selection used as a reference, which remains quiescent following another selection, and the contamination as the fraction of blue galaxies according to the main selection, which become quiescent following another selection. We performed the selection in two redshift bins ($0.5\le z<0.75$ and $0.75\le z\le 0.9$) and the results are shown in Tab.~\ref{tab:comparison_selections}. No selection criterion provides completeness of 100\% together with no contamination. Therefore no selections are equivalent, making comparisons with literature a bit more complicated. When we use the NUVrK diagram to select quiescent galaxies, most of the other selections have good completeness, except for $U-V$ and $n$, which appear to be much more restrictive. The $D_n4000$ appears less restrictive, with most of the NUVrK passive population being selected with $\sim$40\% of blue galaxies that would be added to the sample. This contamination is important for studies of the MSR, as it directly impacts the slope of the passive population, as shown below. We note the very good completeness of the clustering method while having star-forming contamination equivalent to the other criteria. We also provide the completeness and the contamination for other selection criteria if they are used as the main reference criterion.\\

\begin{table*}[h]
\small
\centering
\caption{Comparison of quiescent galaxies selections}             
\centering
\begin{tabular}{|c|c|c|c|c|c|c|c|c|c|c|c|c|c|c|c|}
\hline
 & Redshift &\multicolumn{2}{|c|}{NUVrK} & \multicolumn{2}{|c|}{NUVrJ} & \multicolumn{2}{|c|}{$D_n4000$} & \multicolumn{2}{|c|}{UVJ} & \multicolumn{2}{|c|}{$U-V$} & \multicolumn{2}{|c|}{$n$} & \multicolumn{2}{|c|}{Class}\\
\hline
\multirow{2}{*}{NUVrK} & $0.5\le z < 0.75$ & \multicolumn{2}{c|}{\cellcolor{gray!50}} & 100.0 & 15.9 & 93.6 & 36.4 & 93.2 & 21.8 & 56.4 & 15.0 & 75.2 & 26.1 & 98.3 & 14.0 \\
& $0.75\le z \le 0.9$ & \multicolumn{2}{c|}{\cellcolor{gray!50}} & 100.0 & 22.5 & 92.8 & 40.7 & 91.4 & 20.6 & 68.3 & 18.2 & 74.0 & 29.6 & 98.5 & 25.1 \\
\hline
\multirow{2}{*}{NUVrJ} & $0.5\le z < 0.75$ & 78.9 & 0.0 & \multicolumn{2}{c|}{\cellcolor{gray!50}} & 90.4 & 27.7 & 89.9 & 11.1 & 53.8 & 9.6 & 92.9 & 43.7 & 87.9 & 7.2 \\
& $0.75\le z \le 0.9$ & 78.5 & 0.0 & \multicolumn{2}{c|}{\cellcolor{gray!50}} & 88.7 & 29.7 & 85.0 & 8.4 & 62.0 & 11.0 & 91.8 & 49.7 & 91.0 & 11.5 \\
\hline
\multirow{2}{*}{$D_n4000$} & $0.5\le z < 0.75$ & 60.2 & 5.4 & 73.6 & 10.7 & \multicolumn{2}{c|}{\cellcolor{gray!50}} & 74.0 & 13.0 & 48.1 & 6.5 & 87.4 & 39.1 & 72.1 & 10.8 \\
& $0.75\le z \le 0.9$ & 65.2 & 8.7 & 79.3 & 17.4 & \multicolumn{2}{c|}{\cellcolor{gray!50}} & 73.0 & 15.3 & 57.3 & 10.9 & 86.3 & 51.3 & 79.9 & 20.8 \\
\hline
\multirow{2}{*}{UVJ} & $0.5\le z < 0.75$ & 71.5 & 4.6 & 87.7 & 8.7 & 88.7 & 27.1 & \multicolumn{2}{c|}{\cellcolor{gray!50}} & 53.8 & 7.8 & 91.6 & 43.3 & 79.1 & 14.2 \\
& $0.75\le z \le 0.9$ & 78.5 & 8.1 & 93.0 & 18.0 & 89.4 & 36.4 & \multicolumn{2}{c|}{\cellcolor{gray!50}} & 67.2 & 11.1 & 90.9 & 54.5 & 86.2 & 27.6 \\
\hline
\multirow{2}{*}{$U-V$} & $0.5\le z < 0.75$ & 70.2 & 21.9 & 83.4 & 30.1 & 91.6 & 40.6 & 84.9 & 30.7 & \multicolumn{2}{c|}{\cellcolor{gray!50}} & 89.1 & 55.8 & 84.7 & 30.9 \\
& $0.75\le z \le 0.9$ & 75.2 & 23.4 & 87.7 & 34.9 & 90.8 & 44.1 & 85.9 & 27.7 & \multicolumn{2}{c|}{\cellcolor{gray!50}} & 88.8 & 62.7 & 91.7 & 38.1 \\
\hline
\multirow{2}{*}{$n$} & $0.5\le z < 0.75$ & 49.6 & 5.6 & 62.5 & 10.1 & 71.6 & 21.8 & 63.5 & 12.5 & 36.9 & 10.3 & \multicolumn{2}{c|}{\cellcolor{gray!50}} & 59.4 & 13.1 \\
& $0.75\le z \le 0.9$ & 55.9 & 9.8 & 69.4 & 16.7 & 72.4 & 31.0 & 62.7 & 16.5 & 44.8 & 16.9 & \multicolumn{2}{c|}{\cellcolor{gray!50}} & 68.9 & 22.8 \\
\hline
\multirow{2}{*}{Class} & $0.5\le z < 0.75$ & 82.5 & 1.5 & 92.4 & 11.5 & 91.6 & 29.1 & 85.3 & 19.7 & 54.8 & 9.2 & 92.2 & 48.7 & \multicolumn{2}{c|}{\cellcolor{gray!50}} \\
& $0.75\le z \le 0.9$ & 79.4 & 2.1 & 92.0 & 12.4 & 88.5 & 28.2 & 80.2 & 17.5 & 61.3 & 7.5 & 90.8 & 54.0 & \multicolumn{2}{c|}{\cellcolor{gray!50}} \\
\hline
\end{tabular}
\tablefoot{Lines indicate the main selection method, while columns show the method we compared against. Each column is separated in two: the completeness and the contamination (in \%) for two redshift bins.}
\label{tab:comparison_selections} 
\end{table*}

\begin{figure*}
	\resizebox{\hsize}{!}{\includegraphics{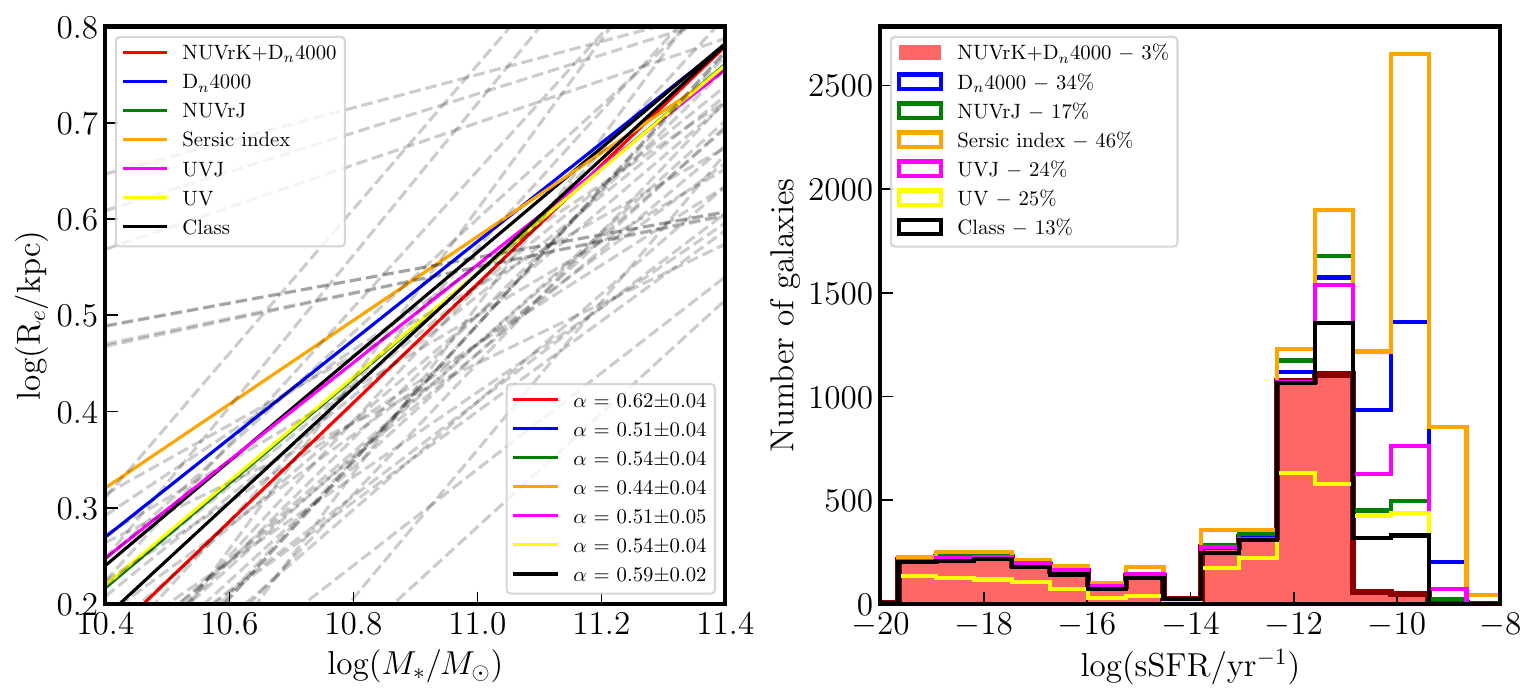}}
	\caption{Comparison of the MSR depending on different selection methods. Left: MSR for the different selection criteria discussed in the text with their associated slope and uncertainty. The dashed gray lines represent the MSRs from the literature shown in Fig.~\ref{fig:mass_size}. Right: sSFR distribution for the different selection criteria with the fraction of blue galaxies (log($\mathrm{sSFR/yr}^{-1})\ge -11$) for each of them shown on the top left corner.}
	\label{fig:sSFR_diffmethods}
\end{figure*}

To check how the slope of the MSR changes with the quiescent selection method, we used the different selection cuts shown in Tab.~\ref{tab:comparison_selections}. Compared to the selection used in this work, the other criteria lead to MSR with milder slopes ranging from 0.44$\pm$0.04 to 0.59$\pm$0.02 (Fig.~\ref{fig:sSFR_diffmethods}, left) and higher intercepts such that the impact of the difference selection mainly affects the low-$M_{*}$ regime. Contamination by the blue population could explain the different slopes as star-forming galaxies are larger with lower stellar mass. In Fig.~\ref{fig:sSFR_diffmethods} (right), we checked the specific star formation rate (sSFR) of the passive sample selected with the different methods. Below a sSFR of $\lesssim -12$~yr$^{-1}$, all the distributions are similar, but a change occurs above this threshold where the sSFR drops for all the selections but the $D_n4000$ and $n$. Taking a threshold between quiescent and star-forming galaxies of log(sSFR/yr$^{-1})=-11$ \citep{sal18}, 46, 34, 25 and 24\% of quiescent galaxies from $D_n4000$, Sersic index, UV and UVJ are found to be active while being lesser than 17\% for all the remaining selections. In particular, the NUVrK+$D_n4000$ selection leads to only 3\% of active galaxies in the final sample. Above this contamination of 24\%, the slope of the MSR is found to be milder and distinct (at 1$\sigma$ level) from the slope derived in this work (Sect.~\ref{Sect:MSR}). These different selections partially explain the different slopes estimated in the literature (shown as gray dashed lines in Fig.~\ref{fig:sSFR_diffmethods}).

\subsection{How different density limits and quiescent selection criteria impacts the size variation in LD and HD environment}

In this work, we used the density contrast (Eq.~\ref{eq:density_contrast}) as a proxy for the environment and defined the HD and LD samples using the density distribution. However, the definition of the environment, and consequently the HD and LD samples, may differ from one study to another. In general HD and LD environments are defined as voids and fields, and groups and clusters, respectively. In \citet{cuc17}, a comparison is made between the LD and HD environments and the voids, groups, and clusters. The $D_{p,5}$ in LD is similar to that in voids while it is larger than the typical size of groups and clusters, although the highest density tail of the HD environment is populated by galaxies in rich groups. The median number density of the field sample in ATLAS$^{\textrm{3D}}$ and of the Coma cluster is log($\Sigma _{p,3}\textrm{)}=-0.5$ and log\,$\Sigma _{p,3}=2$ \citep{cap13} corresponding to $D_{p,5}=2.89$ and 0.16~Mpc, respectively. The threshold for the HD environment in VIPERS ($D_{p,5}=2.85$~Mpc) is therefore similar to the $D_{p,5}$ estimated with ATLAS$^{\textrm{3D}}$. Therefore the HD environment is less dense compared to works using clusters and groups but cannot be strictly compared to field since it comprises more than half of the rich groups in VIPERS (see Fig.~2 in \citealt{cuc17}).

\begin{figure*}
	\resizebox{\hsize}{!}{\includegraphics{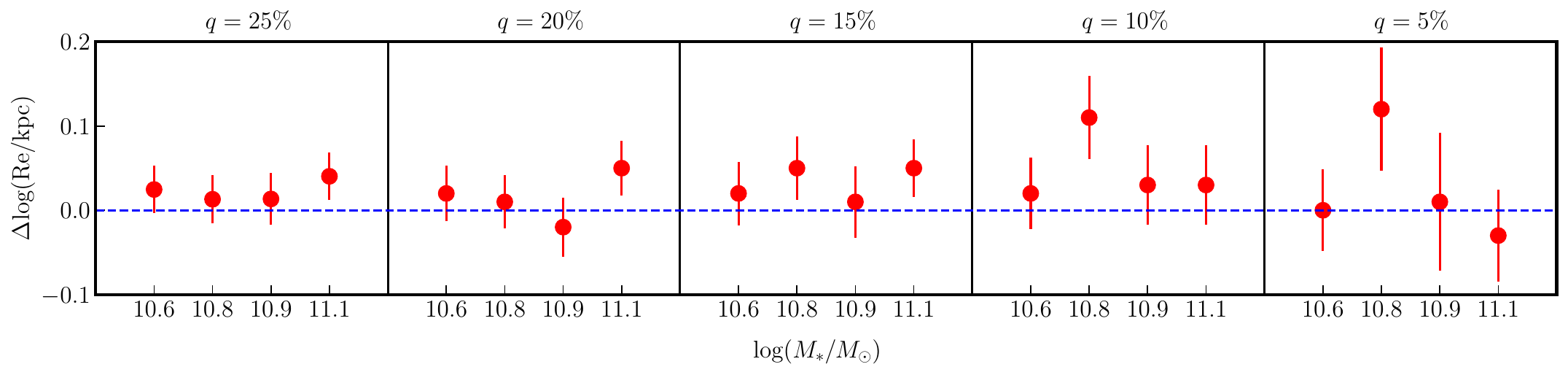}}
	\caption{Size difference between the LD and HD environments (red points) in four mass bins. Each plot is done for different $q$ values representing the thresholds used to define the LD and HD samples from the density distribution.}
	\label{fig:SizeDiff_dens}
\end{figure*}

In Fig.~\ref{fig:MS_MM}, we show that no significant size difference is observed between the LD and HD environments.
It may be that the HD environment traced by VIPERS is not dense enough to highlight an environmental impact on the size. To check if the definition of HD and LD environments impacts the observed size difference, we choose more extreme thresholds to define both density regimes. By reducing the portion of the density distribution considered as HD and LD environments (see Fig.~\ref{fig:red_density_histogram}), we are increasing the probability of finding a difference in the mean size of ETGs. Nonetheless, increasing the density difference between the LD and HD environments comes at the cost of decreasing the statistic of both samples.\\
Instead of defining LD (HD) environment as the first (fourth) quartile ($q=25$\%) of the density distribution, we defined them as the least dense (denser) 20, 15, 10, and 5\% part of the density distribution and recomputed the MSR for the new MM samples. We show in Fig.~\ref{fig:SizeDiff_dens} the size difference, $\Delta \textrm{log(}R_e\textrm{)}=R_e$(HD)-$R_e$(LD), in each stellar mass bin for different definitions of LD and HD samples represented by their $q$ value. The first plot for $q=25\%$ shows the same result than Fig~\ref{fig:MS_MM} in a different way. For $q=20\%$, $15\%$, and $5\%$, the mean size in HD and LD environments in each mass bin agrees within $1-2\sigma$. The second stellar mass bin shows a peculiar behavior at $q=10$\% since the size difference increases but agrees within $3\sigma$. While this difference remains not statistically significant, we do not have explanations for the higher difference in this stellar mass bin.\\

\begin{figure*}
	\resizebox{\hsize}{!}{\includegraphics{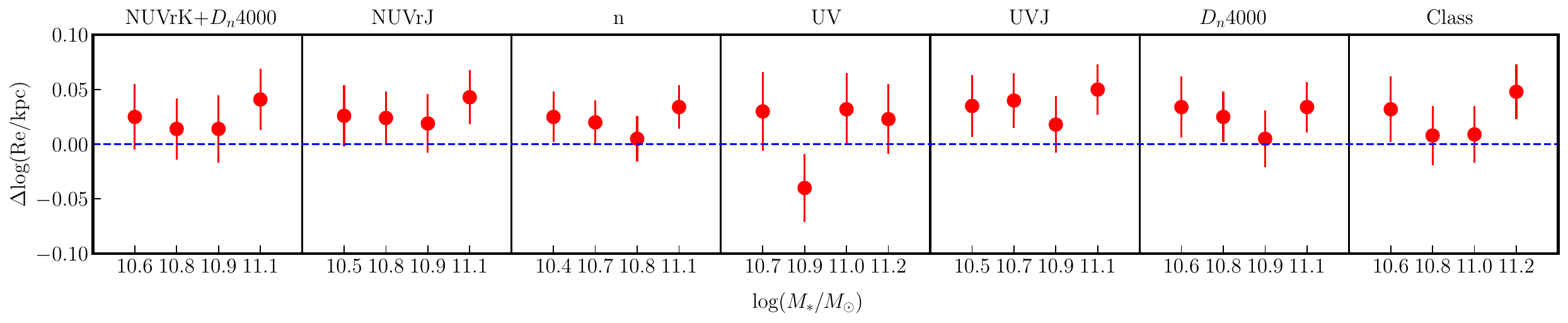}}
	\caption{Size difference between the LD and HD environments (red points) in four mass bins. Each plot is done for different selection criteria: NUVrK+$D_n4000$ (this work), NUVrJ, n, UV, UVJ, and $D_n4000$.}
	\label{fig:SizeDiff_densmethod}
\end{figure*}

In the previous section, we observed that the selection criterion used to define the quiescent population has a mild impact on the slope of the MSR due, at least partially, to the contamination by star-forming galaxies that follow a different MSR. We checked if this selection could change the conclusions about the environmental impact. We reproduced the analysis performed above but with different selection criteria instead of different $q$ values (Fig.~\ref{fig:SizeDiff_densmethod}) and keeping $q=25\%$. We observe that the mean size in the LD and HD environments agrees within $1-2\sigma$ ($3\sigma$ for the fourth mass bin using the UVJ selection). Because no statistical significant size difference is observed between the HD and LD environments, we conclude that the partial contamination by star-forming galaxies that results from different selection criteria is not sufficient to change our conclusions about the environmental impact on the MSR.

\subsection{Growth of galaxies from $z=0.9$ to $z=0.5$}

The mean size of quiescent galaxies evolves with redshift, being more compact at higher $z$ (see Fig.~\ref{fig:mass_size_z}) compared to the local Universe. The present consensus favors the scenario where dry minor mergers are mostly responsible for the size evolution at $z<1$. While major mergers increase the size proportionally to the mass, minor mergers are more efficient in terms of size enhancement. Following the virial theorem \citep{bin08}, the merger size depends on the mass ratio, $\eta$, of both merging galaxies (e.g., \citealt{naa09a,bez09}). For minor mergers ($0.10\le \eta \le 0.25$), the radius increases by a factor $(1+\eta)^2$ and, as a result, the evolutionary path of a galaxy after a minor merger event has a slope $\gamma \sim 2$ in the mass-size plane. Following virial arguments, \citet{new12} express $\gamma$, also called the growth efficiency, as a function of $\alpha$ (the slope of the MSR) and $\eta$:

\begin{equation}\label{eq:size_growth_newman}
    \gamma = 2 - \frac{\textrm{log}(1+\eta^{2-\alpha})}{\textrm{log}(1+\eta)}.
\end{equation}

For $\alpha=0.62$ (Fig.~\ref{fig:mass_size}), and a minor merger scenario ($\eta=1/10$), $\gamma$ is equal to 1.6, in good agreement with \citet{new12}. In other works, $\gamma$ was found to vary for minor mergers with values ranging from 1.3 \citep{nip09}, 1.14 \citep{swe17}, 1.6 \citep{new12}, $\sim 2$ \citep{naa09a,ham22}, 2.24 \citep{oog16} to $\gtrsim 2.7$ \citep{bez09}, and between 1 and 2.5 \citep{bel14b}.\\

To check how the mass and size evolve as the redshift decreases, we divided the sample into two bins $z_1$ and $z_2$. For a consistent comparison, we applied a cut to the sample of galaxies at both redshift ranges at log($M_{*}/M_{\odot})=10.9$, which is the completeness limit at $z=0.9$. The quiescent population at $z_2$ will not be the direct progenitors of the quiescent population at $z_1$ due to the newly quenched galaxies that joined the population between $z_2$ and $z_1$ \citep{car13}. The fraction of newcomers entering the quiescent population depends on the choice of $z_1$ and $z_2$. To avoid a contamination by newcomers, we choose $z_1$ and $z_2$ so that the comoving number density of quiescent galaxies remains constant. In this way, we ensure that we are following the same population from $z_2$ to $z_1$ \citep{mar14}. The sample was cut at $z=0.65$ (whole sample) and $z=0.70$ (LD and HD samples) and we estimated the comoving number density $\phi$ in the two redshift bins. To correct for the selections of VIPERS, each galaxy was weighted by its target sampling rate and spectroscopic success rate (TSR and SSR, see \citealt{gari14}), and the $V_{max}$ method is used:

\begin{align}
&\phi =\sum_{i} \frac{w_i}{V_{max}} = \sum_{i} \frac{(\mathrm{TSR}_i\times \mathrm{SSR}_i)^{-1}}{V_{max}^{i}},\\
&V_{max}^i = \left[V(z_{max}^i) - V(z_{min}^i)\right] \times \frac{\Omega}{4\pi}.
\end{align}

Here $V_{max}^i$ is the maximum volume in which the $i^{th}$ galaxy can be observed as part of the survey, estimated between $z_{min}^i$ and $z_{max}^i$ that were computed based on the absolute luminosities obtained from the SED fitting, and $\Omega$ is the solid angle of the VIPERS surveyed field ($\sim 24$~deg$^{2}$). We found a comoving density of $2.94\times 10^{-5}$, $6.95\times 10^{-6}$, and $4.24\times 10^{-6}$ for the whole sample, the HD and the LD samples, respectively.

\begin{figure*}[ht!]
	\resizebox{0.33\hsize}{!}{\includegraphics{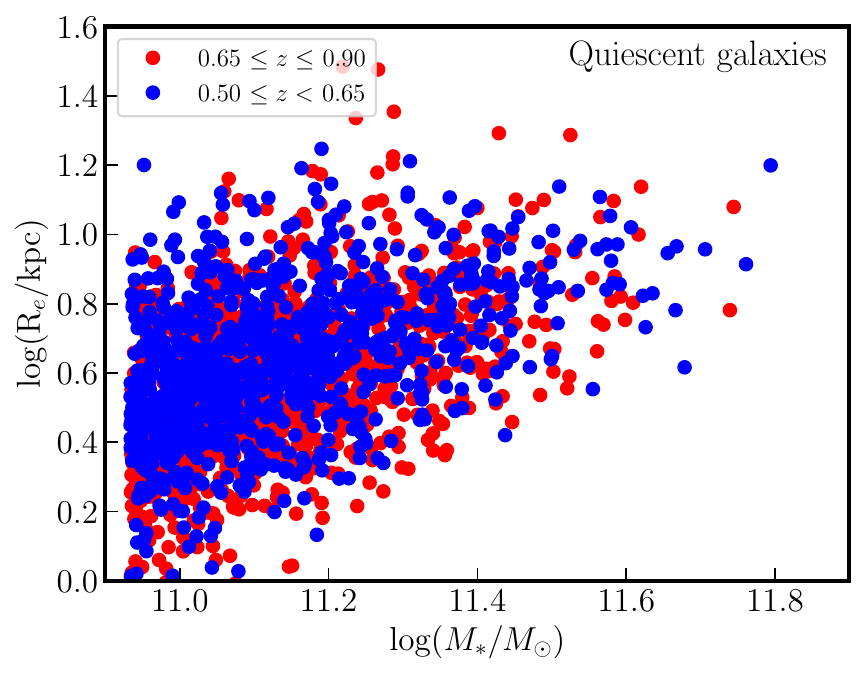}}\resizebox{0.33\hsize}{!}{\includegraphics{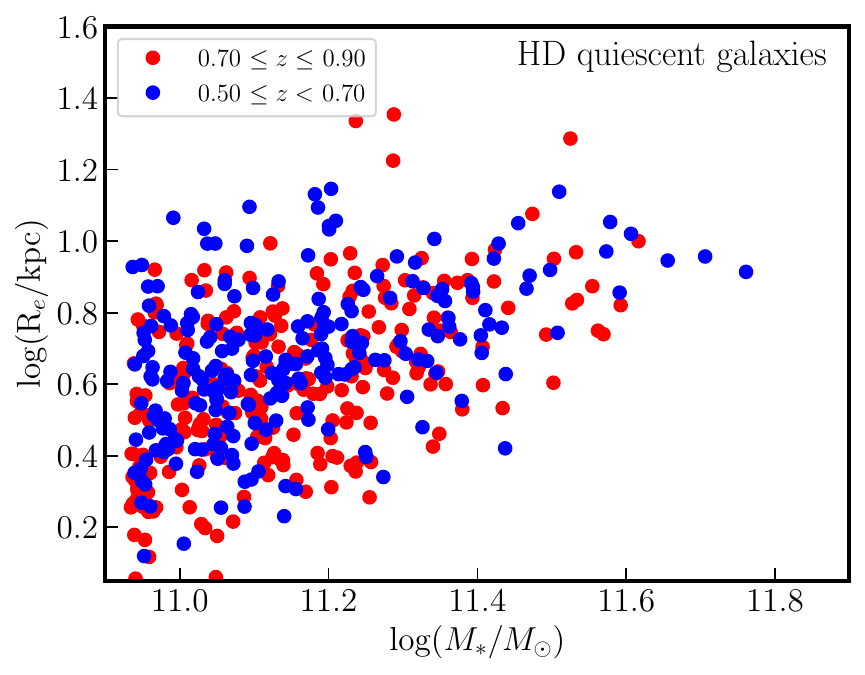}}\resizebox{0.33\hsize}{!}{\includegraphics{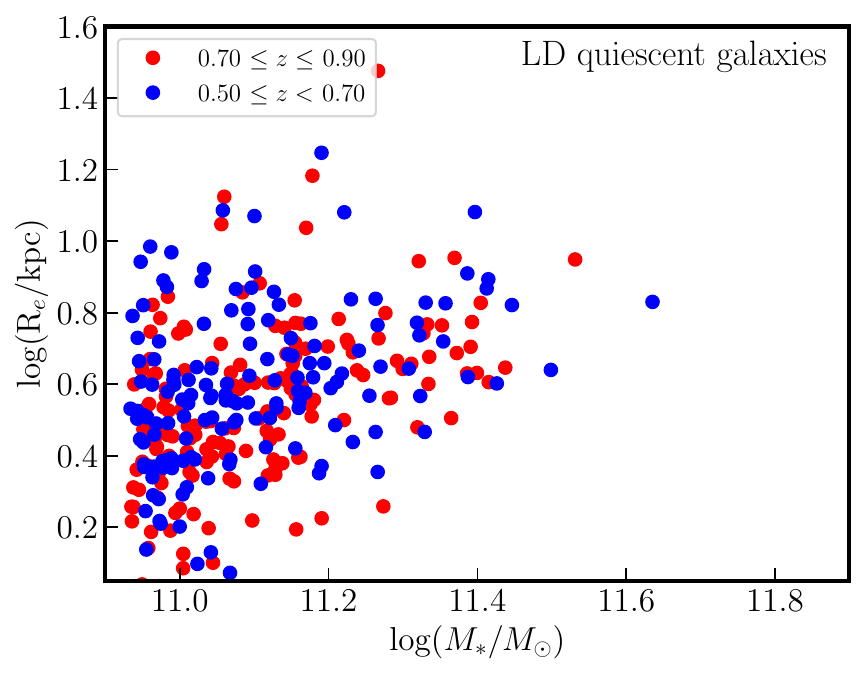}}	
	\caption{MSR for the whole (left), HD (center), and LD (right) samples of passive galaxies in the two redshift bins (lower in blue and higher in red).}
	\label{fig:mass_size_ldhd_2bin}
\end{figure*}

We present in Fig.~\ref{fig:mass_size_ldhd_2bin} the mass-size plane for the whole population, the HD and LD environments. The median size of the population increases from 4.42 to 4.82~kpc (whole sample), from 4.62 to 5.22~kpc (HD), and from 4.08 to 4.44~kpc (LD) while the median stellar mass changes by less than 0.02~dex. In the minor merger scenario, the size and mass of galaxies that merge are related by the relation \citep{naa09b}

\begin{equation}
    \frac{R_f}{R_1}\approx (1+\eta)^2 =  \left(1+\frac{M_{2,*}}{M_{1,*}} \right)^2,
\end{equation}

where $R_1$ is the size of the galaxy, $R_f$ the size of the merged galaxy, $M_{1,*}$ and $M_{2,*}$ the mass of the galaxies that are merging. The change in stellar mass associated to the observed change in size from $0.7\le z\le 0.9$ to $0.5\le z\le 0.7$ is very minor (lesser than 0.03~dex) and would be difficult to detect. This is may be due to the low elapsed time of $\sim 2$~Gyr between both redshift bins. For instance, \citet{ham22} found that the average size evolution from $z=1.1$ to $z=0.7$ is in agreement with a series of 3 minor mergers within 2~Gyr, leading to a difference in stellar mass of 0.13~dex. Given the very low evolution of stellar mass in VIPERS, it is difficult to confirm if $\Delta R\propto \Delta M_*^2$ as it should be for minor mergers. However, the change in size depends on the environment, with an increase of 13\% in the HD environment and an increase of 9\% in the LD environment. This fact in agreement with minor mergers as the merging rate should be lower in LD environments, leading to a smaller size evolution compared to denser environments. This is also in agreement with \citet{yoo17} who found that the merger history of the most massive galaxies ($M_*\ge 10^{11}$) depends on the environment and \citet{coo12} who found that the structural evolution of quiescent galaxies at $0.4\le z \le 1.2$ happens in dense environments.

For the whole, the HD and LD samples, the mean size is increasing by a factor of 1.10, 1.13, and 1.09, corresponding to $\eta$ values ranging from 0.045 to 0.062. These values are lower than the usual range $0.10\le \eta \le 0.25$, indicating that not all the quiescent galaxies have experienced a minor merger event from $z=0.9$ to $z=0.5$. Considering $\eta=0.25$ and 0.10, 17 to 45\% of the population in the LD and 23 to 62\% in the HD environment should experience such merging process to increase the mean size by 9 and 13\% within 2~Gyr. It corresponds to an average merger rate of 0.16 and 0.21~Gyr$^{-1}$ in the LD and HD environments, respectively.\\

How do these merger rates compared with other works from the literature? Using VANDELS and LEGA-C, \citet{ham22} found that the mean size evolution from $z=1.3$ and $z=0.6$ is in agreement with a series of 3 minor mergers with $\eta=0.1$, that is around one minor merger per Gyr. Using the minor merger ($0.10\le \eta \le 0.25$) rate from \cite{lot11}, \citet{dam23} found that minor mergers could explain the mean size evolution from $z=0.6$ to $z=0.2$ in the HectoMAP survey. Other works also found that minor mergers could explain the mean size evolution at $z\le 1.2$ (i.e., \citealt{coo12, dam19, dam22}). From the minor merger rate per galaxy from \citet{lot11} (their tab.~4), the number of minor mergers at $0.5\le z \le 0.9$ ranges between 0.58 to 0.74 depending on the method used to estimate this rate. A comparable minor merger rate of $\sim$0.5~Gyr$^{-1}$ is found by \citet{yoo17} at $z\sim 0.7$ in the lowest dense environment for log($M_*/M_{\odot})\ge 11.2$, leading to one minor merger for the redshift range of this work. The work of \citet{oog13} explains the evolution of the size from $z=2$ to $z=0$ with a sequential minor merger rate every 0.2~Gyr with however a lower mass ratio ($1/20\le \eta \le 1/10$). These values support the number of minor merger required to explain the observe size growth in VIPERS. However we note that other works found much lower values for minor merger rate. \citet{con22} found that 0.14 minor merger occurs over $0.5\le z \le 0.9$, which is too low to explain the mean size evolution in VIPERS, and the same conclusion is reached by \citet{man16} who found that additional processes, in addition to mergers, are needed to explain the strong size evolution from $z=2.5$ downwards and \citet{lopez-s11} in which the upper-limit on the number of minor mergers from $z=0.9$ and $z=0.5$ is equal to 0.13, this last work being however focusing on luminous galaxies. This discrepancy in the merger rate seems to originate, at least partially, from the dataset used as the merger rate in \citet{lot11}, \citet{oog13} and \citet{yoo17} are based on simulations while the ones in \citet{man16}, \citet{lopez-s11}, and \citet{con22} are based on observations.

Based on previous works and on the observed size evolution at a roughly constant stellar mass, the major merger scenario seems to be discarded at $z\le 1$. \citet{ber11} proposed the curvature of the MSR above $2\times 10^{11}$~$M_{\odot}$ to be due to major dry mergers. The last bin of Fig.~\ref{fig:mass_size} is slightly higher than the linear relation, and the dispersion of galaxies above and below this $M_{*}$ could hide such curvature. Instead of a binning per quartile, we divide the sample into 10, 15, and 20 bins with equal numbers of galaxies, to see if any curvature occurs at $M_*\le 2\times 10^{11}$~$M_{\odot}$. For each binning, no curvature is observed, suggesting that major dry mergers are not responsible for the size evolution of the quiescent population at $z\le 1$.\\

Using 2000 massive (log($M_*$/$M_{\odot})\ge 11$) galaxies from VIPERS, \citet{gar17} studied the evolution of low and high stellar mass density galaxies at redshift $0.5\le z \le 1.0$. They found that: 1) the comoving number density of dense galaxies is constant while increasing by a factor of four for less dense objects, 2) less dense galaxies are younger than the dense ones, and 3) the decreasing comoving number density of massive star-forming galaxies is compensated by the increasing comoving number density of quiescent galaxies over the same period of time. They concluded that the main channel by which the quiescent population is built up is the quenching of star-forming galaxies while a build-up through mergers is ruled out. With a sample of 900 massive passive galaxies from VIPERS \citet{gar19} confirm this result but notice that the higher number of less dense galaxies in the HD environment is in agreement with satellite accretion, but only for $\le 1\%$ of low-density massive passive galaxies.\\
Some differences between the sample selection of this work and their can be noted. \citet{gar17,gar19} used the redshift-independent NUVrK colors selection and the physical parameters catalog from \citet{dav16} while we used the redshift-dependent NUVrK selection from \citet{mou16a} and the associated physical parameters catalog based on more photometric bands. In \citet{gar19} the highest density bin is set at log($1+\delta)=0.84$, which is slightly higher than in this work (log($1+\delta)=0.73$) so their HD environment is slightly denser with this limit than our.\\
In the present work we study the size evolution in two bins with an equal comoving number density of galaxies to remove the progenitor bias, in order to look at the residual size evolution that would be caused by other processes. Because we found the size evolution to be environment-dependent, this points toward a scenario where mergers are involved. The associated $\eta$ for the merger are 0.045 and 0.062 for the LD and HD environments and $\eta=0.045$ for the whole sample of massive galaxies.
The similar value for the LD environment and the whole population indicates that the merger rate does not increase with increasing density but rather shows an increase only in the HD environment. This is in line with \citet{gar19} (see their Fig.~8) where the number of low-density passive galaxies is flat in the first three density bins and increases in the last density bin (the HD environment), indicating that the evolution is mostly independent of the environment except at high density where satellite accretion may play a role.\\
From Fig.~\ref{fig:mass_size_z} and Fig.~\ref{fig:MShdld_MM}, the mean size of the quiescent population from the last to the first redshift bin increases by 30\%. With an increase by minor mergers of 9 and 13\% in the LD and HD environments, the mean size increase due to minor merger accounts for 30 and 40\% in the LD and HD environments while the remaining size increase may be due to progenitor bias. Because the size increase for the whole population is similar to that of the LD environment, the minor merger contribution is identical. This result is different from \citet{gar19} where progenitor bias totally accounts for the size evolution in the first three density bins and minor mergers plays a role in the HD environment. Nevertheless we retrieved two general results from \citet{gar19}: 1) the evolution in the first three density bins is independent of environment and 2) minor merger is not the dominant mechanism by which the galaxies evolve in size - the progenitor bias may be more important - and plays a more significant role in the HD environment compared to the first three density bins.\\

\section{Conclusions}\label{Sect:conclusions}

In this paper, we used the VIPERS PDR-2 catalog, a spectroscopic survey of $\sim$90~000 galaxies, to study the MSR of the quiescent population and the impact of the environment on their evolution. The effective circularized radius was computed from $i$-band images using GALFIT through a Sersic profile. The stellar mass and rest-frame photometry were estimated through SED fitting with LEPHARE based on the CFHTLS photometric catalog complemented by GALEX.
The $D_n4000$ spectral break was measured directly from the VIPERS spectra. The environment was traced by the density contrast estimated using a cylinder of 1000~km~s$^{-1}$ half-length and a radius given by the fifth nearest neighbor, where the LD and HD environments were defined as the first and fourth quartile of the density distribution. We selected the quiescent population using the NUVrK diagram and the $D_n4000$ spectral break. The final sample contains 4786 quiescent galaxies and 3998 quiescent galaxies with good density measurements. Our main results are the following:

\begin{itemize}

    \item The slope and intercept of the MSR, $\alpha=0.62\pm 0.04$ and $\textrm{log}(A)=0.52\pm 0.01$, are in agreement with the slopes and intercepts from the literature at a similar redshift. The intercept decreases with increasing redshift, following $R_e(z)=(8.20\pm 0.34)\times (1+z)^{-1.70\pm 0.08}$ and $R_e(z)=(4.76\pm 0.38)\times (1+z)^{-0.81\pm 0.12}$ when using additional values from the literature.\\

    \item Newcomers, defined as quiescent galaxies with $1.5\le D_n{4000}\le 1.6$, are preferentially found in LD environment. In average, newcomers have lower $n$, $M_*$, $R_e$, and higher sSFR when compared to the oldest quiescent galaxies ($D_n{4000}\ge 1.95$).\\
    
    \item We do not find a significant difference ($\le 3\sigma$) in size between the HD and LD environments when using MM samples to suppress the impact of in situ processes depending on the stellar mass. The size between both environments remains identical (i.e., within $3\sigma$) even when changing the definition of LD and HD environments, that is to say when choosing a smaller percentile of the density distribution.\\
    
    \item A positive correlation is observed between $D_n4000$ and the $M_*$ in the HD and LD environments, similar to what was already observed for the whole population. The ages of quiescent galaxies in HD and LD environments are similar within $2\sigma$ at low mass and $1\sigma$ at high mass. The higher offset at low mass is consistent with newcomers being preferentially found in the LD environment.\\

    \item The slope of the MSR slightly increases when using a different quiescent galaxy selection ($D_n4000$, NUVrJ, n, UVJ, and UV) due to a higher fraction of star-forming galaxies, but remains similar within $2\sigma$ and in agreement with the literature.\\

    \item By dividing the sample in two redshift bins with equal comoving number density, we observed that the increase in size is larger in the HD than in the LD environment. Using the model of \citet{naa09a}, we found that 31 and 43\% of the quiescent population in the LD and HD environments experienced a minor merger ($0.10\le \eta \le 0.25$) within $0.5\le z \le 0.9$ at $M\ge 10^{10.9}$~$M_{\odot}$. Using the offset of the MSR between $0.5\le z \le 0.9$, we estimated that minor mergers are responsible for 30 to 40\% of the total size increase of the quiescent population.
    
\end{itemize}

\begin{acknowledgements}
We thank the anonymous referee for his comments that help to improve and clarify the results presented in this work. AP, JK, and MF acknowledge support from the Polish National Science Centre via the grants UMO-2018/30/E/ST9/00082 and UMO-2022/47/D/ST9/00419. MF and MS have been supported by the Polish National Agency for Academic Exchange (Bekker grants BPN/BEK/2023/1/00036/DEC/01 and BPN/BEK/2021/1/00298/DEC/1). MS acknowledges support from the European Union's Horizon 2020 Research and Innovation programme under the Maria Sklodowska-Curie grant agreement (No. 754510).
\end{acknowledgements}

%
   \bibliographystyle{aa} 
   \bibliography{biblio}
   
%

\clearpage
\onecolumn

\begin{appendix} 
\newpage

\newpage

\section{MSR parameters from the literature}

\renewcommand{\arraystretch}{1.1}
{
\scriptsize
\begin{longtable}{l|r|r|r|c|c|c|r}
\caption{MSR parameters of ETGs and properties of the sample used in this work and from the literature. (1) reference, (2) slope, (3) intercept, (4) redshift range, (5) photometric/spectroscopic redshift, (6) number of galaxies ($\sim$ indicates that the number of galaxies was estimated/counted from plots and $<$ indicates that only an upper limit could be found), (7) criteria for quiescent galaxies selection, (8) mass range (in log($M_*/M_{\odot}$)). The MSR parameters, if estimated by eye, have no uncertainties unless the $\pm 1\sigma$ lines were plotted, in which case the uncertainty on log(A) was also estimated by eye.}\\
\label{tab:MS_literature}     
Reference & \multicolumn{1}{|c|}{$\alpha$} & \multicolumn{1}{|c|}{log(A)} & Redshift range & $z$ type & $N_{\textrm{gal}}$ & Selection & Mass range \\
(1)&(2)&(3)&(4)&(5)&(6)&(7)&(8)\\
\hline
\hline
\endfirsthead
\hline\hline
Reference & \multicolumn{1}{|c|}{$\alpha$} & \multicolumn{1}{|c|}{log(A)} & \multicolumn{1}{|c|}{Redshift range} & $z$ type & $N_{\textrm{gal}}$ & Selection & \multicolumn{1}{|c|}{Mass range} \\
(1)&\multicolumn{1}{|c|}{(2)}&\multicolumn{1}{|c|}{(3)}&\multicolumn{1}{|c|}{(4)}&(5)&(6)&(7)&\multicolumn{1}{|c|}{(8)}\\
\hline
\hline
\endhead
This work & $0.62\pm 0.04$ & $0.52\pm 0.01$ & $0.5\le z\le 0.9$ & Spec & 4786 & NUVrK+$D_n4000$ & 9.3 $-$ 11.6 \\
          & $0.64\pm 0.01$ & $0.59\pm 0.01$ & $0.50\le z< 0.58$ & Spec & 1197 & NUVrK+$D_n4000$ & 10.1 $-$ 11.8 \\
          & $0.64\pm 0.04$ & $0.57\pm 0.01$ & $0.58\le z< 0.65$ & Spec & 1196 & NUVrK+$D_n4000$ & 10.3 $-$ 11.8 \\
          & $0.73\pm 0.02$ & $0.52\pm 0.01$ & $0.65\le z< 0.74$ & Spec & 1198 & NUVrK+$D_n4000$ & 10.5 $-$ 11.7 \\          
          & $0.75\pm 0.03$ & $0.47\pm 0.01$ & $0.74\le z\le 0.90$ & Spec & 1195 & NUVrK+$D_n4000$ & 10.7 $-$ 11.7 \\                
HD-MM     & $0.66\pm 0.07$ & $0.54\pm 0.02$ & $0.5\le z\le 0.9$ & Spec & 826 & NUVrK+$D_n4000$ & 10.1 $-$ 11.7 \\
          & $0.63\pm 0.06$ & $0.61\pm 0.02$ & $0.50\le z< 0.58$ & Spec & 183 & NUVrK+$D_n4000$ & 10.1 $-$ 11.4 \\ 
          & $0.65\pm 0.02$ & $0.58\pm 0.01$ & $0.58\le z< 0.65$ & Spec & 182 & NUVrK+$D_n4000$ & 10.4 $-$ 11.4 \\ 
          & $0.72\pm 0.09$ & $0.52\pm 0.02$ & $0.65\le z< 0.74$ & Spec & 206 & NUVrK+$D_n4000$ & 10.5 $-$ 11.5 \\ 
          & $0.84\pm 0.22$ & $0.50\pm 0.03$ & $0.74\le z\le 0.90$ & Spec & 207 & NUVrK+$D_n4000$ & 10.7 $-$ 11.6 \\ 
LD-MM     & $0.62\pm 0.05$ & $0.51\pm 0.01$ & $0.5\le z\le 0.9$ & Spec & 826 & NUVrK+$D_n4000$ & 10.1 $-$ 11.7 \\  
          & $0.62\pm 0.05$ & $0.58\pm 0.02$ & $0.50\le z< 0.58$ & Spec & 183 & NUVrK+$D_n4000$ & 10.1 $-$ 11.4 \\  
          & $0.78\pm 0.05$ & $0.57\pm 0.02$ & $0.58\le z< 0.65$ & Spec & 182 & NUVrK+$D_n4000$ & 10.4 $-$ 11.4 \\  
          & $0.76\pm 0.05$ & $0.48\pm 0.02$ & $0.65\le z< 0.74$ & Spec & 206 & NUVrK+$D_n4000$ & 10.5 $-$ 11.5 \\  
          & $0.81\pm 0.05$ & $0.47\pm 0.02$ & $0.74\le z< 0.90$ & Spec & 207 & NUVrK+$D_n4000$ & 10.7 $-$ 11.5 \\ 
\hline          
\citet{all15}\tablefootmark{$1$} & $0.76\pm 0.04$ & $0.41\pm 0.16$ & $2.0\le z\le 2.2$ & Phot & 7 & UVJ & 9.8 $-$ 11.5 \\
\citet{all15} & $0.76\pm 0.04$ & $0.34\pm 0.09$ & $2.0\le z\le 2.2$ & Phot & 30 & UVJ & 9.5 $-$ 11.2 \\
\hline
\citet{bar22}\tablefootmark{$2$} & $0.54$ & $0.67\pm 0.02$ & $0.014\le z\le 0.1$ & Spec & 524 & MS & 10 $-$ 11.5 \\
\citet{bar22} & $0.66$ & $-1.36\pm 0.02$ & $0.6\le z< 0.68$ & Spec & 219 & MS & 10.5 $-$ 11.5 \\
\citet{bar22} & $0.78$ & $-3.16\pm 0.05$ & $0.68\le z\le 0.76$ & Spec & 273 & MS & 10.5 $-$ 11.5 \\
\hline
\citet{bel15}\tablefootmark{$3$} & $0.76$ & $0.31$ & $1.0\le z\le 1.6$ & Spec & 51 & sSFR & 10.7 $-$ 11.5 \\
\hline
\citet{cha16}\tablefootmark{$4$} & $0.138\pm 0.192$ & $0.40\pm 2.55$ & $z=1.39$ & Spec & 12 & sSFR & 10 $-$ 11.5 \\
\citet{cha16} & $0.447\pm 0.268$ & $0.37\pm 7.29$ & $z=1.39$ & Spec & 12 & sSFR & 10.5 $-$ 11.5 \\
\citet{cha16} & $0.359\pm 0.135$ & $0.44\pm 2.41$ & $z=1.39$ & Phot/Spec & 36 & sSFR & 10 $-$ 11.5 \\
\citet{cha16} & $0.576\pm 0.173$ & $0.40\pm 6.58$ & $z=1.39$ & Phot/Spec & 36 & sSFR & 10.5 $-$ 11.5 \\
\hline
\citet{che22}\tablefootmark{$5$} & $0.69$ & $0.66$ & $0.012\le z\le 0.324$ & Spec & 4437 & MS/Mz relation & 10 $-$ 12 \\
\hline
\citet{cim12}\tablefootmark{$6$} & $0.52\pm 0.05$ & $0.69$ & $0.0\le z< 0.6$ & Phot/Spec & $\sim$360 & Color/Spectra/sSFR/Morph & 10.5 $-$ 11.8 \\
\citet{cim12} & $0.47\pm 0.04$ & $0.4$ & $0.6\le z< 0.9$ & Phot/Spec & $\sim$360 & Color/Spectra/sSFR/Morph & 10.5 $-$ 11.8 \\
\citet{cim12} & $0.50\pm 0.04$ & $0.27$ & $0.9\le z\le 3.0$ & Phot/Spec & $\sim$360 & Color/Spectra/sSFR/Morph & 10.5 $-$ 11.6 \\
\hline
\citet{dam11}\tablefootmark{$7$} & $0.47\pm 0.06$ & $0.42$ & $0.2\le z< 0.8$ & Spec & 212 & Spectro/Morph/Color & 10 $-$ 12 \\
\citet{dam11} & $0.51\pm 0.06$ & $0.41$ & $0.8\le z< 1.4$ & Spec & 199 & Spectro/Morph/Color & 10 $-$ 12.1 \\
\citet{dam11} & $0.52\pm 0.12$ & $0.21$ & $1.4\le z< 2.0$ & Spec & 44 & Spectro/Morph/Color & 10 $-$ 11.8 \\
\citet{dam11} & $0.51\pm 0.36$ & $-0.06$ & $2.0\le z\le 2.7$ & Spec & 10 & Spectro/Morph/Color & 10 $-$ 11.2 \\
\hline
\citet{dam19} & $0.79\pm 0.05$ & $0.65\pm 0.02$ & $0.16\le z< 0.26$ & Spec & 492 & $D_n4000$ & 10.2 $-$ 11.6 \\
\citet{dam19} & $0.90\pm 0.04$ & $0.58\pm 0.02$ & $0.26\le z< 0.36$ & Spec & 1527 & $D_n4000$ & 10.2 $-$ 12.2 \\
\citet{dam19} & $0.87\pm 0.03$ & $0.51\pm 0.01$ & $0.36\le z< 0.48$ & Spec & 840 & $D_n4000$ & 10.6 $-$ 11.9 \\
\citet{dam19} & $0.70\pm 0.09$ & $0.49\pm 0.03$ & $0.48\le z\le 0.65$ & Spec & 487 & $D_n4000$ & 11 $-$ 12.1 \\
\hline
\citet{dam22} & $0.510\pm 0.012$ & $0.76\pm 0.01$ & $0.05\le z\le 0.07$ & Spec & 31001 & $D_n4000$ & 10 $-$ 11.7 \\
\citet{dam22} & $0.643\pm 0.024$ & $0.61\pm 0.01$ & $0.1\le z\le 0.6$ & Spec & 2906 & $D_n4000$ & 9.2 $-$ 12.3 \\
\hline
\citet{dam23}\tablefootmark{$8$} & $0.67\pm 0.01$ & $0.57\pm 0.01$ & $0.2\le z\le 0.5$ & Spec & 23113 & $D_n4000$ & 10 $-$ 12 \\
\citet{dam23} & $0.661\pm 0.009$ & $0.799\pm 0.003$ & $0.2\le z< 0.3$ & Spec & 9205 & $D_n4000$ & 10.4 $-$ 11.8 \\
\citet{dam23} & $0.749\pm 0.011$ & $0.762\pm 0.020$ & $0.3\le z< 0.4$ & Spec & 7710 & $D_n4000$ & 10.6 $-$ 11.9 \\
\citet{dam23} & $0.788\pm 0.014$ & $0.720\pm 0.003$ & $0.4\le z< 0.5$ & Spec & 6198 & $D_n4000$ & 10.6 $-$ 11.9 \\
\citet{dam23} & $0.882\pm 0.029$ & $0.614\pm 0.010$ & $0.5\le z\le 0.6$ & Spec & 2949 & $D_n4000$ & 10.9 $-$ 12 \\
\hline
\citet{dela14}\tablefootmark{$9$} & $0.52\pm 0.08$ & $0.52\pm 0.03$ & $0.7\le z< 0.9$ & Phot/Spec & 130 & Morph & 10.5 $-$ 11.6 \\
\citet{dela14} & $0.48\pm 0.08$ & $0.48\pm 0.03$ & $0.9\le z< 1.1$ & Phot/Spec & 96 & Morph & 10.5 $-$ 11.6 \\
\citet{dela14} & $0.34\pm 0.10$ & $0.44\pm 0.04$ & $1.1\le z\le 1.6$ & Phot/Spec & 94 & Morph & 10.5 $-$ 11.6 \\
\citet{dela14} & $0.47\pm 0.07$ & $0.47\pm 0.02$ & $0.7\le z< 0.9$ & Phot/Spec & 123 & Morph & 10.5 $-$ 11.6 \\
\citet{dela14} & $0.57\pm 0.07$ & $0.47\pm 0.02$ & $0.9\le z< 1.1$ & Phot/Spec & 135 & Morph & 10.5 $-$ 11.6 \\
\citet{dela14} & $0.5\pm 0.1$ & $0.30\pm 0.02$ & $1.1\le z\le 1.6$ & Phot/Spec & 125 & Morph & 10.5 $-$ 11.6 \\
\hline
\citet{dia19d}\tablefootmark{$10$} & $0.71\pm 0.02$ & $0.73\pm 0.04$ & $0.1\le z< 0.3$ & Phot & $\sim$50 & MCDE & 10.5 $-$ 11.6 \\
\citet{dia19d} & $0.71\pm 0.02$ & $0.65\pm 0.04$ & $0.3\le z< 0.5$ & Phot & $\sim$150 & MCDE & 10.5 $-$ 11.5 \\
\citet{dia19d} & $0.71\pm 0.02$ & $0.57\pm 0.05$ & $0.5\le z< 0.7$ & Phot & $\sim$180 & MCDE & 10.5 $-$ 11.5 \\
\citet{dia19d} & $0.71\pm 0.02$ & $0.49\pm 0.05$ & $0.7\le z\le 0.9$ & Phot & 263 & MCDE & 10.5 $-$ 11.6 \\
\hline
\citet{fai17}\tablefootmark{$11$} & $0.55\pm 0.05$ & $0.5$ & $0.5\le z< 1.0$ & Phot & $\sim$800 & NUVrJ & 10 $-$ 12 \\
\citet{fai17} & $0.62\pm 0.09$ & $0.3$ & $1.0\le z< 1.5$ & Phot & $\sim$800 & NUVrJ & 10 $-$ 11.8 \\
\citet{fai17} & $0.59\pm 0.15$ & $0.18$ & $1.5\le z\le 2.0$ & Phot & $\sim$750 & NUVrJ & 10 $-$ 11.9 \\
\hline
\citet{fav18}\tablefootmark{$12$} & $0.238\pm 0.044$ & $0.67\pm 0.18$ & $0.2\le z< 0.3$ & Spec & <75441 & De Veaucouleurs profile & 11.1 $-$ 12 \\
\citet{fav18} & $0.219\pm 0.022$ & $0.70\pm 0.11$ & $0.3\le z< 0.43$ & Spec & <75441 & De Veaucouleurs profile & 11.1 $-$ 12 \\
\citet{fav18} & $0.202\pm 0.021$ & $0.73\pm 0.12$ & $0.43\le z< 0.55$ & Spec & <153304 & De Veaucouleurs profile & 11 $-$ 12.1 \\
\citet{fav18} & $0.172\pm 0.015$ & $0.75\pm 0.12$ & $0.55\le z\le 0.6$ & Spec & <153304 & De Veaucouleurs profile & 11.1 $-$ 12.2 \\
\hline
\citet{fern13}\tablefootmark{$13$} & $0.56$ & $0.62$ & $0.01\le z\le 0.05$ & Spec & 15 & Morph/$n$ & 9.9 $-$ 11.4 \\
\citet{fern13} & $0.56$ & $0.58$ & $0.01\le z\le 0.05$ & Spec & 67 & Morph & 9.4 $-$ 11.4 \\
\citet{fern13} & $0.54$ & $0.54\pm 0.01$ & $0.01\le z\le 0.05$ & Spec & 23 & Morph/$n$ & 10.1 $-$ 11.3 \\
\citet{fern13} & $0.60$ & $0.66\pm 0.01$ & $0.01\le z\le 0.05$ & Spec & $\sim$500 & Morph/$n$ & 9.3 $-$ 11.6 \\
\hline
\citet{gar17} & $0.59\pm 0.07$ & $0.60\pm 0.01$ & $0.5\le z< 0.7$ & Spec & 782 & NUVrK & 11 $-$ 11.5 \\
\citet{gar17} & $0.70\pm 0.08$ & $0.53\pm 0.02$ & $0.7\le z< 0.9$ & Spec & 868 & NUVrK & 11 $-$ 11.5 \\
\citet{gar17} & $0.52\pm 0.10$ & $0.53\pm 0.02$ & $0.9\le z\le 1.0$ & Spec & 372 & NUVrK & 11 $-$ 11.5 \\
\hline
\citet{guo09}\tablefootmark{$14$} & $0.91\pm 0.03$ & $0.08$ & $0.0\le z\le 0.08$ & Spec & 911 & CEN/$n$ & 9.9 $-$ 11.7 \\
\citet{guo09} & $0.70\pm 0.05$ & $0.63$ & $0.0\le z\le 0.08$ & Spec & $\sim$450 & CEN/$n$ & 9.9 $-$ 11.7 \\
\hline
\citet{ham22} & $0.56\pm 0.04$ & $0.47\pm 0.02$ & $0.6\le z\le 0.8$ & Spec & 377 & UVJ & 10.3 $-$ 11.5 \\
\citet{ham22} & $0.72\pm 0.06$ & $0.28\pm 0.03$ & $1.0\le z\le 1.3$ & Spec & 137 & UVJ & 10.3 $-$ 11.6 \\
\hline
\citet{hon23}\tablefootmark{$15$} & $0.88$ & $0.53$ & $D\le 110$~Mpc &  & 202 & Bulge/Spheroid & 9.5 $-$ 12 \\
\hline
\citet{hue13b}\tablefootmark{$16$} & $0.59\pm 0.09$ & $0.59\pm 0.10$ & $0.2\le z< 0.5$ & Phot/Spec & 59 & NUVr & 10.5 $-$ 11.4 \\
\citet{hue13b} & $0.50\pm 0.11$ & $0.34\pm 0.08$ & $0.5\le z< 0.8$ & Phot/Spec & 123 & NUVr & 10.5 $-$ 11.7 \\
\citet{hue13b} & $0.59\pm 0.05$ & $0.28\pm 0.02$ & $0.8\le z\le 1.0$ & Phot/Spec & 210 & NUVr & 10.5 $-$ 11.8 \\
\citet{hue13b} & $0.52\pm 0.03$ & $0.47\pm 0.03$ & $0.2\le z< 0.5$ & Phot/Spec & 128 & NUVr & 10.5 $-$ 11.9 \\
\citet{hue13b} & $0.56\pm 0.04$ & $0.41\pm 0.03$ & $0.5\le z< 0.8$ & Phot/Spec & 110 & NUVr & 10.5 $-$ 11.7 \\
\citet{hue13b} & $0.49\pm 0.04$ & $0.41\pm 0.03$ & $0.8\le z\le 1.0$ & Phot/Spec & 155 & NUVr & 10.5 $-$ 11.9 \\
\hline
\citet{ich12}\tablefootmark{$17$} & $0.126\pm 0.009$ & $0.54\pm 0.01$ & $0.25\le z\le 3.0$ & Phot/Spec & 408 & UVJ & 8 $-$ 11.5 \\
\citet{ich12} & $0.129\pm 0.02$ & $0.54\pm 0.03$ & $0.25\le z\le 0.5$ & Phot/Spec & 28 & UVJ & 8 $-$ 11.5 \\
\citet{ich12} & $0.133\pm 0.018$ & $0.55\pm 0.02$ & $0.5\le z\le 0.75$ & Phot/Spec & 70 & UVJ & 8.2 $-$ 11.5 \\
\citet{ich12} & $0.118\pm 0.017$ & $0.56\pm 0.01$ & $0.75\le z\le 1.0$ & Phot/Spec & 134 & UVJ & 9 $-$ 11.5 \\
\citet{ich12} & $0.153\pm 0.024$ & $0.55\pm 0.01$ & $1.0\le z\le 1.25$ & Phot/Spec & 83 & UVJ & 9.5 $-$ 11.5 \\
\citet{ich12} & $0.125\pm 0.037$ & $0.57\pm 0.02$ & $1.25\le z\le 1.5$ & Phot/Spec & 32 & UVJ & 9.5 $-$ 11.5 \\
\citet{ich12} & $0.166\pm 0.074$ & $0.52\pm 0.03$ & $1.5\le z\le 2.0$ & Phot/Spec & 28 & UVJ & 10.3 $-$ 11.5 \\
\citet{ich12} & $0.250\pm 0.083$ & $0.48\pm 0.02$ & $2.0\le z\le 2.5$ & Phot/Spec & 27 & UVJ & 10.3 $-$ 11.5 \\
\citet{ich12} & $0.48\pm 0.07$ & $0.48\pm 0.02$ & $2.5\le z\le 3.0$ & Phot/Spec & 6 & UVJ & 10.5 $-$ 11.2 \\
\citet{ich12} & $0.132\pm 0.008$ & $0.54\pm 0.01$ & $0.25\le z\le 3.0$ & Phot/Spec & 445 & UVJ & 8 $-$ 11.5 \\
\citet{ich12} & $0.165\pm 0.023$ & $0.55\pm 0.03$ & $0.25\le z\le 0.50$ & Phot/Spec & 31 & UVJ & 8 $-$ 11.5 \\
\citet{ich12} & $0.137\pm 0.017$ & $0.55\pm 0.02$ & $0.50\le z\le 0.75$ & Phot/Spec & 73 & UVJ & 8.2 $-$ 11.5 \\
\citet{ich12} & $0.119\pm 0.015$ & $0.56\pm 0.01$ & $0.75\le z\le 1.0$ & Phot/Spec & 141 & UVJ & 9 $-$ 11.5 \\
\citet{ich12} & $0.150\pm 0.022$ & $0.55\pm 0.01$ & $1.0\le z\le 1.25$ & Phot/Spec & 88 & UVJ & 9.5 $-$ 11.5 \\
\citet{ich12} & $0.152\pm 0.045$ & $0.56\pm 0.02$ & $1.25\le z\le 1.5$ & Phot/Spec & 35 & UVJ & 9.5 $-$ 11.5 \\
\citet{ich12} & $0.077\pm 0.069$ & $0.48\pm 0.02$ & $1.5\le z\le 2.0$ & Phot/Spec & 36 & UVJ & 10.3 $-$ 11.5 \\
\citet{ich12} & $0.281\pm 0.068$ & $0.48\pm 0.02$ & $2.0\le z\le 2.5$ & Phot/Spec & 32 & UVJ & 10.3 $-$ 11.5 \\
\citet{ich12} & $0.251\pm 0.133$ & $0.43\pm 0.04$ & $2.5\le z\le 3.0$ & Phot/Spec & 9 & UVJ & 10.3 $-$ 11.2 \\
\hline
\citet{kaw21} & $0.34\pm 0.01$ & $0.55$ & $0.2\le z< 0.4$ & Phot & 40259 & urz & 8.5 $-$ 11.8 \\
\citet{kaw21} & $0.36\pm 0.01$ & $0.53$ & $0.4\le z< 0.6$ & Phot & 82714 & urz & 8.5 $-$ 12 \\
\citet{kaw21} & $0.41\pm 0.01$ & $0.47$ & $0.6\le z< 0.8$ & Phot & 61486 & urz & 8.5 $-$ 12 \\
\citet{kaw21} & $0.44\pm 0.01$ & $0.42$ & $0.8\le z\le 1.0$ & Phot & 57059 & urz & 8.8 $-$ 12.1 \\
\hline
\citet{kro14} & $0.82\pm 0.22$ & $0.30\pm 0.08$ & $1.85\le z\le 2.3$ & Phot/Spec & 34 & UVJ & 10.6 $-$ 11.7 \\
\citet{kro14} & $0.53\pm 0.29$ & $0.29\pm 0.07$ & $1.85\le z\le 2.3$ & Spec & 14 & UVJ & 10.8 $-$ 11.7 \\
\hline
\citet{kuc17}\tablefootmark{$18$} & $0.43$ & $0.55$ & $z=0.44$ & Spec & $\sim$293 & BRI & 9.2 $-$ 11.3 \\
\hline
\citet{lan13} & $0.31$ & $0.45$ & $0.5\le z< 1.0$ & Phot & $\sim$2900 & UVJ/sSFR & 9.8 $-$ 12.2 \\
\citet{lan13} & $0.44$ & $0.3$ & $1.0\le z\le 2.0$ & Phot & $\sim$2200 & UVJ/sSFR & 10.4 $-$ 11.9 \\
\hline
\citet{lan15}\tablefootmark{$19$} & $0.63\pm 0.03$ & $0.580\pm 0.003$ & $0.01\le z\le 0.1$ & Spec & 1300 & Morph & 10.3 $-$ 11.3 \\
\hline
\citet{lange16}\tablefootmark{$20$} & $0.329\pm 0.010$ & $0.51\pm 0.01$ & $0.002\le z\le 0.06$ & Spec & 806 & Morph & 8 $-$ 11.2 \\
\citet{lange16} & $0.643\pm 0.032$ & $0.64\pm 0.09$ & $0.002\le z\le 0.06$ & Spec & $\sim$400 & Morph & 10 $-$ 11.2 \\
\citet{lange16} & $0.786\pm 0.048$ & $0.64\pm 56.99$ & $0.002\le z\le 0.06$ & Spec & $\sim$400 & Morph & 10.3 $-$ 11.2 \\
\hline
\citet{mal10}\tablefootmark{$21$} & $0.26\pm 0.07$ & $0.47\pm 0.06$ & $0.122\le z\le 0.205$ & Phot & 167 & Hubble type & 9 $-$ 11.5 \\
\citet{mal10} & $0.30\pm 0.04$ & $0.55\pm 0.04$ & $0.05\le z\le 0.3$ & Phot & 89 & Hubble type & 9 $-$ 11.5 \\
\hline
\citet{mcl13}\tablefootmark{$22$} & $0.56$ & $0.23$ & $1.3\le z\le 1.5$ & Spec & 41 & sSFR & 10.8 $-$ 11.7 \\
\citet{mcl13} & $0.56$ & $0.24$ & $1.3\le z\le 1.5$ & Spec & 37 & $n$ & 10.8 $-$ 11.7 \\
\citet{mcl13} & $0.56$ & $0.24$ & $1.3\le z\le 1.5$ & Spec & <41 & Formation time (old) & 10.8 $-$ 11.7 \\
\citet{mcl13} & $0.56$ & $0.24$ & $1.3\le z\le 1.5$ & Spec & <41 & Formation time (young) & 10.8 $-$ 11.7 \\
\citet{mcl13} & $0.56$ & $0.23$ & $1.3\le z\le 1.5$ & Spec & <37 & $n$/Formation time (old) & 10.8 $-$ 11.7 \\
\citet{mcl13} & $0.56$ & $0.26$ & $1.3\le z\le 1.5$ & Spec & <37 & $n$/Formation time (young) & 10.8 $-$ 11.7 \\
\hline
\citet{mil23}\tablefootmark{$23$} & $0.62\pm 0.05$ & $0.40\pm 0.01$ & $1.0\le z< 1.2$ & Phot & 279 & UVJ & 10.3 $-$ 11.4 \\
\citet{mil23} & $0.57\pm 0.05$ & $0.32\pm 0.01$ & $1.2\le z< 1.4$ & Phot & 252 & UVJ & 10.3 $-$ 11.4 \\
\citet{mil23} & $0.44\pm 0.07$ & $0.30\pm 0.02$ & $1.4\le z< 1.6$ & Phot & 184 & UVJ & 10.3 $-$ 11.4 \\
\citet{mil23} & $0.51\pm 0.06$ & $0.28\pm 0.02$ & $1.6\le z< 1.8$ & Phot & 343 & UVJ & 10.3 $-$ 11.4 \\
\citet{mil23} & $0.55\pm 0.10$ & $0.26\pm 0.03$ & $1.8\le z\le 2.0$ & Phot & 205 & UVJ & 10.3 $-$ 11.4 \\
\hline
\citet{mos20}\tablefootmark{$24$} & $0.680\pm 0.053$ & $0.45\pm 0.01$ & $0.3\le z< 0.7$ & Phot/Spec & $\sim$303 & UVJ & 10.2 $-$ 11.4 \\
\citet{mos20} & $0.770\pm 0.016$ & $0.36\pm 0.01$ & $0.7\le z< 1.0$ & Phot/Spec & $\sim$303 & UVJ & 10.6 $-$ 11.4 \\
\citet{mos20} & $0.81\pm 0.14$ & $0.320\pm 0.004$ & $1.0\le z< 1.3$ & Phot/Spec & $\sim$605 & UVJ & 10.5 $-$ 11.4 \\
\citet{mos20} & $1.08\pm 2.24$ & $0.220\pm 0.002$ & $1.3\le z\le 2.0$ & Phot/Spec & $\sim$152 & UVJ & 10.9 $-$ 11.5 \\
\hline
\citet{mow19a} & $0.48\pm 0.03$ & $0.60\pm 0.02$ & $0.0\le z< 0.5$ & Phot & <788 & UVJ & 9 $-$ 12 \\
\citet{mow19a} & $0.58\pm 0.04$ & $0.47\pm 0.02$ & $0.5\le z< 1.0$ & Phot & <788 & UVJ & 9 $-$ 12 \\
\citet{mow19a} & $0.73\pm 0.02$ & $0.33\pm 0.01$ & $1.0\le z< 1.5$ & Phot & <788 & UVJ & 9 $-$ 12 \\
\citet{mow19a} & $0.63\pm 0.05$ & $0.21\pm 0.02$ & $1.5\le z< 2.0$ & Phot & <203 & UVJ & 9 $-$ 12 \\
\citet{mow19a} & $0.48\pm 0.12$ & $0.06\pm 0.06$ & $2.0\le z< 2.5$ & Phot & <203 & UVJ & 9 $-$ 12 \\
\citet{mow19a} & $0.59\pm 0.23$ & $0.15\pm 0.11$ & $2.5\le z\le 3.0$ & Phot & <203 & UVJ & 9 $-$ 12 \\
\hline
\citet{nad21} & $0.38\pm 0.03$ & $0.59\pm 0.05$ & $0.0\le z\le 2.0$ & Phot & 122 & SED & 6.0 $-$ 11.1 \\
\citet{nad21} & $0.37\pm 0.07$ & $0.51\pm 0.01$ & $0.0\le z< 0.5$ & Phot & 37 & SED & 6.0 $-$ 10.8 \\
\citet{nad21} & $0.38\pm 0.06$ & $0.50\pm 0.08$ & $0.5\le z< 1.0$ & Phot & 68 & SED & 8.5 $-$ 11.1 \\
\citet{nad21} & $0.56\pm 0.32$ & $0.59\pm 0.36$ & $1.0\le z\le 2.0$ & Phot & 17 & SED & 10 $-$ 10.9 \\
\hline
\citet{ned21} & $0.68\pm 0.04$ & $0.67\pm 0.01$ & $0.2\le z< 0.5$ & Phot/Spec & 253 & UVJ & 10.3 $-$ 11.4 \\
\citet{ned21} & $0.64\pm 0.03$ & $0.50\pm 0.01$ & $0.5\le z< 1.0$ & Phot/Spec & 539 & UVJ & 10.3 $-$ 11.7 \\
\citet{ned21} & $0.63\pm 0.04$ & $0.32\pm 0.01$ & $1.0\le z< 1.5$ & Phot/Spec & 430 & UVJ & 10.3 $-$ 11.6 \\
\citet{ned21} & $0.61\pm 0.05$ & $0.22\pm 0.01$ & $1.5\le z\le 2.0$ & Phot/Spec & 469 & UVJ & 10.3 $-$ 11.7 \\
\hline
\citet{new12}\tablefootmark{$25$} & $0.59\pm 0.07$ & $0.46\pm 0.02$ & $0.4\le z< 1.0$ & Phot & $\sim$193 & sSFR/no 24~$\mu$m & 10.7 $-$ 11.9 \\
\citet{new12} & $0.62\pm 0.09$ & $0.30\pm 0.02$ & $1.0\le z< 1.5$ & Phot & $\sim$139 & sSFR/no 24~$\mu$m & 10.7 $-$ 11.9 \\
\citet{new12} & $0.63\pm 0.11$ & $0.21\pm 0.02$ & $1.5\le z< 2.0$ & Phot & $\sim$108 & sSFR/no 24~$\mu$m & 10.7 $-$ 11.7 \\
\citet{new12} & $0.69\pm 0.17$ & $0.04\pm 0.04$ & $2.0\le z\le 2.5$ & Phot & $\sim$43 & sSFR/no 24~$\mu$m & 10.7 $-$ 11.5 \\
\hline
\citet{new14}\tablefootmark{$26$} & $0.61\pm 0.07$ & $0.07\pm 0.06$ & $1.7\le z\le 1.9$ & Phot & $\sim$200 & UVJ & 10.7 $-$ 11.9 \\
\hline
\citet{sar09}\tablefootmark{$27$} & $1.19\pm 0.47$ & $0.47\pm 0.09$ & $1.015\le z\le 1.921$ & Spec & 32 & Spectra & 10 $-$ 12 \\
\hline
\citet{sar11}\tablefootmark{$28$} & $1.10\pm 0.72$ & $0.70\pm 0.28$ & $0.964\le z\le 1.921$ & Spec & 62 & Spectra/Morph/$n$ & 9.77 $-$ 11.8 \\
\hline
\citet{sar14}\tablefootmark{$29$} & $1.92\pm 0.99$ & $0.91\pm 0.34$ & $z=1.27$ & Phot/Spec & 16 & Morph & 9.7 $-$ 11.3 \\
\hline
\citet{sar17}\tablefootmark{$30$} & $0.50\pm 0.06$ & $0.50\pm 0.07$ & $1.0\le z\le 1.45$ & Phot/Spec & 489 & Morph & 10.5 $-$ 21 \\
\hline
\citet{she03} & $0.56$ & $0.62$ & $0.05\le z\le 0.15$ & Spec & $\sim$35923 & $n$ & 10 $-$ 12 \\
\hline
\citet{sue19}\tablefootmark{$31$} & $0.4077\pm 0.3890$ & $0.33\pm 0.25$ & $1.0\le z< 1.5$ & Phot/Spec & $\sim$250 & UVJ & 10.5 $-$ 12 \\
\citet{sue19} & $0.757\pm 0.37$ & $0.71\pm 0.49$ & $1.5\le z< 2.0$ & Phot/Spec & $\sim$200 & UVJ & 10.5 $-$ 12 \\
\citet{sue19} & $0.775\pm 1.02$ & $0.09\pm 0.11$ & $2.0\le z\le 2.5$ & Phot/Spec & $\sim$100 & UVJ & 10.5 $-$ 12 \\
\hline
\citet{swe17}\tablefootmark{$32$} & $0.74\pm 0.06$ & $0.49\pm 0.04$ & $z=1.067$ & Phot/Spec & 49 & $n$ & 9.9 $-$ 11.7 \\
\citet{swe17} & $0.45\pm 0.04$ & $0.53\pm 0.05$ & $z=1.067$ & Phot/Spec & 48 & $n$ & 10 $-$ 11.9 \\
\citet{swe17} & $0.44\pm 0.08$ & $0.46\pm 0.05$ & $z=1.067$ & Phot/Spec & 43 & $n$ & 9.9 $-$ 11.9 \\
\citet{swe17} & $0.84\pm 0.06$ & $0.48\pm 0.03$ & $z=1.067$ & Phot/Spec & $\sim$48 & $n$ & 9.9 $-$ 11.9 \\
\citet{swe17} & $1.77\pm 0.21$ & $0.15\pm 0.02$ & $z=1.067$ & Phot/Spec & $\sim$26 & $n$ & 9.9 $-$ 11.9 \\
\hline
\citet{tof12} & $0.59$ & $0.07$ & $1.8\le z\le 2.2$ & Spec & 4 & Post-starburst/SFR & 11 $-$ 11.7 \\
\hline
\citet{van14}\tablefootmark{$33$} & $0.75\pm 0.06$ & $0.68\pm 0.02$ & $0.0\le z< 0.5$ & Phot/Spec & $\sim$500 & UVJ & 9 $-$ 11.3 \\
\citet{van14} & $0.71\pm 0.03$ & $0.49\pm 0.01$ & $0.5\le z< 1.0$ & Phot/Spec & $\sim$2540 & UVJ & 9 $-$ 12 \\
\citet{van14} & $0.76\pm 0.04$ & $0.30\pm 0.01$ & $1.0\le z< 1.5$ & Phot/Spec & $\sim$1400 & UVJ & 9 $-$ 11.6 \\
\citet{van14} & $0.76\pm 0.04$ & $0.17\pm 0.02$ & $1.5\le z< 2.0$ & Phot/Spec & $\sim$1400 & UVJ & 9.3 $-$ 11.8 \\
\citet{van14} & $0.76\pm 0.04$ & $0.03\pm 0.01$ & $2.0\le z< 2.5$ & Phot/Spec & $\sim$300 & UVJ & 9.7 $-$ 11.8 \\
\citet{van14} & $0.79\pm 0.07$ & $0.03\pm 0.02$ & $2.5\le z\le 3.0$ & Phot/Spec & $\sim$650 & UVJ & 10 $-$ 11.3 \\
\hline
\citet{wat22} & $0.51$ & $0.59$ & Nearby galaxies & & 126 & Morph & 10.7 $-$ 11.5 \\
\hline
\citet{wil10}\tablefootmark{$34$} & $0.54\pm 0.06$ & $0.46\pm 0.02$ & $0.5\le z< 1.0$ & Phot & & sSFR & 10.6 $-$ 11.6 \\
\citet{wil10} & $0.56\pm 0.06$ & $0.35\pm 0.01$ & $1.0\le z< 1.5$ & Phot & & sSFR & 10.6 $-$ 11.6 \\
\citet{wil10} & $0.50\pm 0.07$ & $0.25\pm 0.01$ & $1.5\le z\le 2.0$ & Phot & & sSFR & 10.6 $-$ 11.6 \\
\hline
\citet{yan21}\tablefootmark{$35$} & $0.76$ & $0.37\pm 0.11$ & $1.0\le z\le 1.5$ & Phot/Spec & $\sim$17 & UVJ & 10.3 $-$ 11.2 \\
\hline
\citet{yoo17}\tablefootmark{$36$} & $0.621$ & $0.66$ & $0.1\le z\le 0.15$ & Spec & $\sim$55000 & urgi/concentration/Morph & 10.7 $-$ 11.2 \\
\citet{yoo17} & $0.851$ & $0.59$ & $0.1\le z\le 0.15$ & Spec & $\sim$18000 & urgi/concentration/Morph & 11.3 $-$ 11.9 \\
\hline
\citet{zan16}\tablefootmark{$37$} & $0.62$ & $0.34$ & $1.05\le z< 1.7$ & Spec & 22 & sSFR & 10.7 $-$ 11.8 \\
\citet{zan16} & $0.62$ & $0.17$ & $1.7\le z\le 2.05$ & Spec & 10 & sSFR & 10.7 $-$ 11.7 \\
\hline
\end{longtable}
\tablefoot{
\tablefoottext{$1$}{ The slope is fixed and taken from \citet{van14}. The relations are for cluster and field galaxies, respectively. }\\
\tablefoottext{$2$}{ ETGs defined as galaxies with $\textrm{SFR}<\textrm{SFR}_{MS}-2\sigma _{RMS}$ with the MS from \citet{whi12}. The MSRs are estimated by eye. }\\
\tablefoottext{$3$}{ The MSR is estimated by eye. }\\
\tablefoottext{$4$}{ MSR for the cluster XMMUJ2235-2557. MSRs using $r_{\textrm{mass}}$ are also available (see their Tab.~2). }\\
\tablefoottext{$5$}{ Quiescent galaxies are defined such as $\textrm{log(SFR)}<0.64\times \textrm{log(}M_*\textrm{)}-7.22$ and further reduced using the $M_*-z$ plane to match the number of post-starburst galaxies.The MSR is estimated by eye. }\\
\tablefoottext{$6$}{ The sample is taken from different works (see their Tab.~1). The intercept is estimated by eye. }\\
\tablefoottext{$7$}{ Work based on different spectroscopic surveys where the quiescent galaxies selection are either spectroscopically selected objects with old stellar population, morphology or color. The intercept is estimated by eye. }\\
\tablefoottext{$8$}{ In addition, MSRs at fixed $D_n4000$ are given (see their Tab.~2) as well as MSRs for newcomers and the aging population (see their Tab.~3). }\\
\tablefoottext{$9$}{ MSRs with free slope for cluster and field galaxies, respectively. MSRs with fixed slope ($\alpha=0.57$) and for each cluster are available (see their Tabs.~7 and 8).}\\
\tablefoottext{$10$}{ MCDE: Rest-frame stellar mass color diagram corrected for extinction (see \citealt{dia19b}). MSRs for galaxies whose properties were derived using the BC03 simple stellar populations models. MSRs using EMILES (\citealt{gir00} Padova00 and \citealt{pie04} BaSTI) are available (see their Tab.~3). }\\
\tablefoottext{$11$}{ The low-mass sample contains 9000 star-forming and quiescent galaxies, and the high-mass samples contains 403 star-forming and quiescent galaxies. The intercept is estimated by eye. }\\
\tablefoottext{$12$}{ The number of red galaxies per redshift range is not indicated. }\\
\tablefoottext{$13$}{ The \citet{she03} slope is used for the two first MSRs, where the Sersic's size and Sextractor's size are used for the two other MSRs. For the last MSR, there are 824 ETGs (Sersic index) but no number are given for the morphological separation. }\\
\tablefoottext{$14$}{ CEN: Central galaxies. The second MSR excludes the bright galaxies ($M_r-5\textrm{log(}h\textrm{)}<-22$). The intercept is estimated by eye. }\\
\tablefoottext{$15$}{ MSR for bulge/spheroid (including LTGs) from different datasets \citep{sav16,dav19,sah19,hon22}. }\\
\tablefoottext{$16$}{ The first three MSRs are for field galaxies and the three last for group galaxies. More MSRs with different ETGs selections are available (see their Tab.~1). }\\
\tablefoottext{$17$}{ MSRs for resolved and resolved+unresvolved galaxies. MSRs using $R_{90}$ are also available (see their Tab.~1). }\\
\tablefoottext{$18$}{ MSR for the cluster MACS J1206.2-0847. The parameters are estimated by eye. There are 543 star-forming and quiescent galaxies in the sample. }\\
\tablefoottext{$19$}{ There are 2010 elliptical galaxies but the sample is cut at $2\times 10^{10}$~$M_{\odot}$ and we take the $g$-band as it is closer to the $i$-band at $z\sim 0.7$ (see their Appendix B). MSRs for ETGs based on different ETGs selection are shown in their Tab.~3. }\\
\tablefoottext{$20$}{ The different MSRs correspond to different range of stellar mass (see their Tab.~1 and their conclusions). }\\
\tablefoottext{$21$}{ The MSR parameters are estimated based on their Table.~3. The relations are for cluster and field galaxies, respectively. The MSR for core ellipticals agrees with cluster elliptical within uncertainties and is not shown here. }\\
\tablefoottext{$22$}{ The formation time is derived using SED fitting. }\\
\tablefoottext{$23$}{ MSRs using $r_{mass}$ are available (see their Tab.~2). }\\
\tablefoottext{$24$}{ There are 1363 quiescent galaxies in the sample. The slope and intercept were retrieved for the high-mass end considering 1+($M_*/M_p)\approx (M_*/M_p$), leading to $\alpha=\beta$ and log(A)=log($r_p$)-$\beta$log($M_p$)+($\beta-\alpha/\delta$)log(1/2) (see their Eq.~3). MSRs for $R_{20}$ and $R_{90}$ are also available (see their Tab.~1). }\\
\tablefoottext{$25$}{ The radius is estimated such as $R_h$=a(1+(b/a))/2. Quiescent galaxies are characterized by $\textrm{sSFR}<0.02$~Gyr$^{-1}$ and no MIPS 24~$\mu$m detection. }\\
\tablefoottext{$26$}{ The MSR is parametrized by the usual linear fit but with an additional dependence on redshift through an additional term ($-0.26(z-1.8)$). }\\
\tablefoottext{$27$}{ The MSR parameters are estimated based on their Table 2. }\\
\tablefoottext{$28$}{ The MSR parameters are estimated based on their Table 1. }\\
\tablefoottext{$29$}{ MSR for the cluster RDCS J0848+4453. The MSR parameters are estimated based on their Tables 3 and 4. }\\
\tablefoottext{$30$}{ MSR obtained using a least-square fit with cluster and field galaxies, which is consistent with an orthogonal fit within 1$\sigma$. The MSR is better fitted with a broken power-law over the entire mass range with $R_e=26\times M_{*}^{-0.13\pm 0.2}$ and $R_e=2.77\times 10^{-7} M_{*}^{-0.64\pm 0.09}$ below and above $2.5\times 10^{10}$~$M_{\odot}$.}\\
\tablefoottext{$31$}{ The MSR is obtained by combining log$(r_{\textrm{mass}}/r_{\textrm{light}})\propto \textrm{log(}M_*)$ and log$(r_{\textrm{mass}}/r_{\textrm{light}})\propto \textrm{log(}r_{light})$. }\\
\tablefoottext{$32$}{ MSR for the cluster SPT-CLJ0546-5345. There is no ETG selection but the median Sersic index is in agreement with ETGs ($n=3.8 \pm 0.5$). }\\
\tablefoottext{$33$}{ The sample (LTGs+ETGs) contains 9130 ($0\le z \le 1$), 16639 ($1< z \le 2$), and 5189 ($2< z \le 3$) galaxies. }\\
\tablefoottext{$34$}{ Quiescent galaxies have $\textrm{sSFR}<0.3/t_H$ where $t_H$ is the age of the Universe at redshift $z$. The number of galaxies is not indicated. }\\
\tablefoottext{$35$}{ The slope is fixed from \citet{van14}. We choose the MSR derived using the Bradac lens model, which is similar within uncertainties to MSR using other lens models. }\\
\tablefoottext{$36$}{ ETGs are defined so that $c<0.43$, red $u-r$ color, slightly negative $\Delta (g-i)$, and refined with a visual classification (see \citealt{par05,cho10}). There are 73116 ETGs in the sample. }\\
\tablefoottext{$37$}{ The MSR is estimated by eye. }\\
}
} 

\end{appendix}

\end{document}